\documentclass[fleqn,usenatbib]{mnras}
\usepackage{newtxtext,newtxmath}

\usepackage[T1]{fontenc}

\DeclareRobustCommand{\VAN}[3]{#2}
\let\VANthebibliography\thebibliography
\def\thebibliography{\DeclareRobustCommand{\VAN}[3]{##3}\VANthebibliography}

\usepackage{graphicx}	
\usepackage{subfigure}
\usepackage{amsmath}	

\title[X-ray halo]{Prospect of Detecting X-Ray Halos Around Middle-Aged Pulsars with eROSITA}

\author[B. Li et al.]{
Ben Li,$^{1,2}$\thanks{E-mail: bli@smail.nju.edu.cn}
Yi Zhang,$^{1,2}$\thanks{E-mail: yi\_zhang@smail.nju.edu.cn}
Ruo-Yu Liu$^{\href{https://orcid.org/0000-0003-1576-0961}{\includegraphics[scale=0.04]{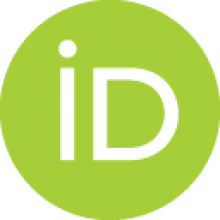}}}$$^{1,2}$\thanks{E-mail: ryliu@nju.edu.cn}
and Xiang-Yu Wang$^{\href{https://orcid.org/0000-0002-5881-335X}{\includegraphics[scale=0.04]{orcid-ID.png}}}$$^{1,2}$\thanks{E-mail: xywang@nju.edu.cn}
\\
$^{1}$School of Astronomy and Space Science, Nanjing University, Nanjing 210023, Jiangsu, China\\
$^{2}$Key laboratory of Modern Astronomy and Astrophysics(Nanjing University), Ministry of Education, Nanjing 210023, People's Republic of China\\
}

\date{Accepted XXX. Received YYY; in original form ZZZ}

\pubyear{2021}

\begin{document}
\label{firstpage}
\pagerange{\pageref{firstpage}--\pageref{lastpage}}
\maketitle

\begin{abstract}
The detection of extended TeV $\gamma$-ray emission (dubbed ``TeV halos'') from Geminga and Monogem pulsars by HAWC collaboration implies that the halo-like morphologies around middle-aged pulsars may be common.
The $\gamma$-rays above 10\,TeV are thought to arise from inverse Compton (IC) scattering of relativistic electrons/positrons in the pulsar halos off cosmic microwave background photons. In the meanwhile, these electrons and positrons can produce X-ray synchrotron emission in the interstellar magnetic field, resulting in a diffuse emission in the X-ray band (namely X-ray halos).
Here, we study the prospect of detecting X-ray halos with eROSITA from 10 middle-aged pulsars with characteristic age $\tau_c$ larger than tens of thousands of  years in the ATNF pulsar catalog. Assuming a benchmark value (i.e., $B=3 \, \mu {\rm G}$) for the magnetic field, most of the X-ray halos are found to be bright enough to be detectable by eROSITA  in the energy range of $\rm 0.5 - 2 \, keV$ during its four-year all-sky survey. Among these pulsar halos, three are supposed to produce X-ray fluxes above the eROSITA sensitivity of the first all-sky survey. Given the good angular resolution and the large field of view, eROSITA is expected to be able to measure the spatial distribution of the X-ray halos from sub-pc scale up to tens of pc scale. The intensity profiles of the X-ray halos are very useful to constrain the magnetic field and the energy-dependence of the diffusion coefficient in the pulsar halos.
\end{abstract}

\begin{keywords}
X-rays: ISM -- ISM: magnetic fields -- pulsars: general
\end{keywords}



\section{Introduction} \label{sec:intro}

It has long been believed that the pulsar wind nebula (PWN) is among the major accelerators of cosmic ray (CR) electrons and positrons. Electrons (hereafter we do not distinguish positrons from electrons unless stated otherwise) accelerated at the termination shock will eventually escape the PWNe and then diffuse in the interstellar medium (ISM) after tens of thousands of years since the pulsar's birth \citep{2020A&A...636A.113G}. It has been predicted that escaped electrons will emit $\gamma$-rays by IC scattering off background photons and form extended gamma-ray sources \citep{2004vhec.book.....A}. Such a prediction has been confirmed by the High Altitude Water Cherenkov observatory (HAWC), which recently detected extended $\gamma$-ray emission ($\rm \sim 30 \, pc$) above a few TeV around two nearby middle-aged pulsars: Geminga and Monogem \citep{2017Sci...358..911A}, as well as by the discovery of the extended source LHAASO J0621+3755 \citep{2021PhRvL.126x1103A} by the Large High Altitude Air Shower Observatory (LHAASO). These halo-like emissions are suggested to constitute a new morphological source class: pulsar TeV halos \citep{2017PhRvD..96j3016L}. 

$\gamma$-rays at energies of 10 TeV are produced by electrons at energies around 60 TeV through IC scattering of the cosmic microwave background (CMB). These electrons could also produce keV photons through synchrotron process in the interstellar magnetic field of typical strength of $3-6\,\mu$G, so a diffuse X-ray emission with a similar morphology to that in the multi-tens-TeV band is expected. However, X-ray halos around middle-aged pulsars have not yet been detected by the current X-ray instruments such as Chandra and XMM-Newton, although diffuse nonthermal X-ray emission has been detected around some PWNe such as the Boomerang Nebula \citep{2021Innov...200118G} and the PWN around PSR~J1825-1334 \citep{2011ApJ...742...62V}. This is probably due to their limited fields of view (e.g., only $16^{\prime} \times 16^{\prime}$ for Chandra) and small number of known pulsar TeV halos.

The extended Roentgen Survey with an Imaging Telescope Array (eROSITA) telescope is one of the two X-ray telescopes (eROSITA and ART-XC) on board the Russian-German space mission Spectrum-Roentgen-Gamma (SRG) \citep{2021arXiv210413267S}. eROSITA is sensitive to X-rays in the energy range of $0.3-11 \rm \, keV$, with a wide field of view of $\rm 0.81 \, deg^2$ and an angular resolution of $18^{\prime \prime}$, aiming at mapping the high-energy Universe in X-rays during its four year all-sky survey. The expected sensitivities achieved by eROSITA for extended sources in the energy range of $0.5-2 \rm \, keV$ are $1.1 \times 10^{-13} \rm \, erg \, cm^{-2} \, s^{-1}$ for the first all-sky survey (eRASS:1), and $3.4 \times 10^{-14} \rm \, erg \, cm^{-2} \, s^{-1}$ for the four year all-sky survey (eRASS:8). The latter is about 20 times more sensitive than that of the ROSAT all sky survey \citep[see][]{2012arXiv1209.3114M}. With an integration of the ``unlimited field of view'' and excellent sensitivity, eROSITA is ideal for detecting faint extended objects like pulsar X-ray halos.

The magnetic field in the pulsar halos could significantly affect the X-ray flux because the power of synchrotron emission is proportional to the energy density of the magnetic field ($U_{\rm B} = B^2 / 8 \pi$). In a recent study, \citet{2019ApJ...875..149L} put a constraint on the magnetic field of the TeV halo of Geminga which is $B <1\, \mu$G, based on the analysis of the XMM-Newton data and Chandra data towards the central $10^{\prime}$ of the Geminga's TeV halo. We will show that eROSITA is able to constrain the magnetic field in the halo at a larger spatial scale based on the measurement of the diffuse X-ray emission around Geminga with its four-year all-sky survey data, with either detection or null detection.  \citet{2017Sci...358..911A} found a low diffusion coefficient $D({\rm 100 \, TeV}) = 4.5 \times 10^{27}  \rm \, cm^2 \, s^{-1}$ in the local environment when jointly fitting the surface brightness profile of Geminga and Monogem, which is about two orders of magnitude smaller than that derived from the boron-to-carbon ratio and its origin is unknown. The radial profile measurement of the X-ray halo by the eROSITA telescope will give us a better understanding of the diffusion coefficient.

The paper is organized as follows. In section \ref{sec:model}, we introduce the model for calculating the particle transport and photon emission. In section \ref{sec:results}, we show that the eROSITA telescope is capable of detecting extended X-ray emission from a number of middle-aged pulsars, and the potential to constrain the magnetic field and diffusion coefficient. A short discussion and summary are presented in section \ref{sec:discussion and summary}.


\section{Model} \label{sec:model}

In this section, we introduce the model for calculating the electron transport and radiation in the pulsar halos. To obtain the flux of IC and synchrotron radiation, we first need to know the spatial and energy distributions of electrons under the joint play of injection, diffusion and cooling. For simplicity, we only consider isotropic diffusion of electrons with a spatially homogeneous diffusion coefficient as well as a homogeneous magnetic field.

\subsection{Electron distribution} \label{subsec:Edis}

Electrons carried by the pulsar wind are accelerated to relativistic energies at the termination shock \citep{2006ARA&A..44...17G}. 
They will eventually diffuse into the ISM and produce a halo around middle-aged pulsars \citep{2020A&A...636A.113G}.
To obtain the electron distribution, we consider the case of continuous injection of electrons from a point source. The energy released into electrons can be expressed by:
\begin{equation}
	W_e(t)=\eta L_{\rm spin}(t)=\eta \dot E_{\rm spin} \left(\frac{1+t/ \tau_0}{1+t_{\rm age}/ \tau_0}\right)^{\rm - \frac{(n+1)}{(n-1)}},
\end{equation}
where $L_{\rm spin}(t)$ is the pulsar spin-down luminosity at time $t$ since pulsar's birth, $\dot E_{\rm spin}$ is the spin-down luminosity at present time, $\eta$ represents the energy conversion efficiency from the pulsar spin-down energy to the electron energy and $\tau_0$ is the initial spin-down timescale of the pulsar (here we set $\tau_0=12$ kyr \footnote{Note that the initial spin-down timescale of pulsars is uncertain. However, since X-ray emitting electrons have very short cooling timescales, the X-ray flux is insensitive to the pulsar injection history.}). We take $n=3$ for the braking index, which assumes the spindown of pulsars is caused by magnetic dipole radiations.

We assume that electrons are injected with a power-law spectrum with an exponential cutoff, i.e., 
\begin{equation}
	Q_e(E_e,t)=Q_0(t) \, E_e^{-p} \, \exp\left(- \frac{E_e}{E_{e,{\rm max}}}\right),
\end{equation}
where $E_{e,{\rm max}}$ is the cutoff energy and $p$ is the spectral index. $Q_0(t)$ is a normalization factor, which can be determined by $W_e(t)=\int_{E_{e, \rm min}}^{E_{e, \rm max}} Q_e(E_e,t) E_e \, dE_e$. We take $E_{e, \rm min}=1 \, \rm GeV$.

Electrons will suffer energy losses when they diffuse in the ISM. Their differential number density $N_e(t,E_e,\vec{r})$ then can be obtained by solving the particle transport equation:
\begin{equation} \label{equation:3}
	\frac{\partial N_e}{\partial t}=D(E_e) \Delta N_e + \frac{\partial}{\partial E_e} (\dot E_e N_e) + Q_e \, \delta(\vec{r}),
\end{equation}
where $D(E_e)$ is the energy dependent diffusion coefficient, which has the form $D(E_e)=D_0 \, (E_e / 100 \, {\rm TeV})^{\delta}$, with $D_0$ being the normalization factor at $\rm 100 \, TeV$ and $\delta=1/3$ for the Kolmogorov turbulence model. $\dot E_e$ is the electron energy loss rate due to IC and synchrotron radiation, which can be expressed by:
\begin{equation}
	\dot E_e=- \frac{4}{3} c \sigma_{\rm T} \gamma^2 \left[ \frac{B^2}{8\pi}+\sum \frac{U_i}{(1+4 \gamma \epsilon_i)^{3/2}} \right],
\end{equation}
where $\sigma_{\rm T}$ is the Thomson cross section and $\gamma$ is the electron's Lorentz factor. $B^2/8\pi$ and $U_i$ are energy densities of interstellar magnetic field and the $i \rm \, th$ component of the interstellar radiation field (ISRF), respectively. The ISRF considered here includes three components: cosmic microwave background (CMB), infrared radiation field and visible light radiation field. For blackbody/greybody radiation with temperature $T$, $\epsilon_i=2.8 k_{\rm B} T / m_e c^2$.

An analytical solution to equation (\ref{equation:3}) at present time $t_{\rm age}$ is given by \citet{1995PhRvD..52.3265A}:
\begin{equation}
	N_e(E_e,r)=\int _{t_{\rm lower}} ^{t_{\rm age}} \frac{\dot E_e (E_e^{\prime})}{\dot E_e (E_e)} \, \frac{Q_e(E_e^{\prime}, t^{\prime})}{\pi^{3/2} \, r_d^3} \, \exp\left(-\frac{r^2}{r_d^2}\right) \, dt^{\prime},
\end{equation}
where $t_{\rm lower}={\rm max}[0,t_{\rm age}+\int _{E_e}^{E_{e,{\rm max}}} \frac{dE_e}{\dot E_e}]$. $E_e^{\prime}$ is the electron energy at time $t^{\prime}$ and satisfies $t_{\rm age}-t^{\prime}= - \int _{E_e}^{E_e^{\prime}} \frac{dE_e}{\dot E_e}$. $r_d$ is the diffusion length scale, given by
\begin{equation}
	r_d=2 \left[ \int _{E_e}^{{\rm min}[E_e^{\prime},E_{e,\rm max}]} \frac{D(E_e^{\prime \prime})}{\dot E_e (E_e^{\prime \prime})} \, dE_e^{\prime \prime} \right]^{1/2}.
\end{equation}

\subsection{Radiation}

Relativistic electrons produce photon emission via \hyphenation{synchro-tron}synchrotron radiation in the interstellar magnetic field and inverse Compton scattering off interstellar background photons. Following the method given by \citet{2019ApJ...875..149L}, the predicted intensity of synchrotron radiation at an angular distance $\theta$ from the pulsar's direction can be calculated by integrating over the contribution of electrons along the line of sight of that direction:
\begin{equation}
	I_{\rm syn}(\epsilon,\theta) = \frac{1}{4\pi}\int \mathcal{F} _{\rm syn} \{N_e[E_e,r(\theta,l)], \ B\}dl,
\end{equation}
where $\mathcal{F} _{\rm syn}$ is an operator calculating the differential spectrum of the synchrotron radiation by electrons in the element volume $N_e(E_e)$ given a magnetic field $B$, and $l$ is the distance of the element volume and Earth.

In analogy, the intensity of the IC radiation, $I_{\rm IC}$, is given by
\begin{equation}
	I_{\rm IC}(\epsilon,\theta) = \frac{1}{4\pi}\int \mathcal{F} _{\rm IC} \{N_e[E_e,r(\theta, l)], \  n_{\rm ph} \}dl,
\end{equation}
where $n_{\rm ph}$ is the differential number density of the interstellar photon field. We neglect the Galactic absorption in the calculation for simplicity \citep{2000ApJ...542..914W}. The total flux can be obtained by integrating over the solid angle around the pulsar.


\section{Results} \label{sec:results}

Electrons accelerated by the PWNe are injected into the ISM after a few tens of kyr since the pulsars' birth, so young pulsars are unlikely to show halo-like emission. On the other hand, since the spin-down luminosity of pulsars continuously decrease  with increasing age, diffuse X-ray emission around old pulsars may be too faint to be detected. Therefore, X-ray halos are only expected to be seen around middle-aged pulsars. For this reason, we search for pulsars with characteristic ages between $\rm 50 - 500 \, kyr$ from the Australia Telescope National Facility (ATNF) pulsar catalog \citep{2005AJ....129.1993M}. Considering the electrons producing X-ray halo also emit TeV photons, we choose pulsars with TeV counterparts in the 3HWC catalog as X-ray halo source candidates. Table \ref{tab:1} shows the top ten pulsars in the rank according to their energy flux at Earth (i.e., $\dot E_{\rm spin} / d^2$, $d$ is distance of pulsar to Earth), and their nearest counterparts in the 3HWC catalog \citep{2020ApJ...905...76A}.

\subsection{SEDs} \label{subsec:SEDs}

In this section, we use the model presented in section \ref{sec:model} to calculate the IC and synchrotron emission for the pulsars listed in Table~\ref{tab:1}. We employ the following values of magnetic field and diffusion coefficient: $B = \rm 3 \, \mu G$ and $D_0 = 3.5 \times 10^{27}  \rm \, cm^2 \, s^{-1}$, which are consistent with that used by \citet{2017Sci...358..911A}.

The resulting SEDs for the pulsar halo candidates are shown in Figure~\ref{fig:1}. The black solid line is the IC flux integrated over a radius of $\rm 60 \, pc$ from the pulsar, which is fitted to the  HAWC data \citep[cyan bow tie, see][]{2017Sci...358..911A, 2020ApJ...905...76A}. The black dashed line represents the synchrotron flux within an angular radius of $30^{\prime}$. For comparison, we also plot the eROSITA sensitivities for extended source of $30^{\prime}$ radius, which are calculated following the method given by \citet{2012arXiv1209.3114M}. For a given extraction region, we first estimate the minimal net counts ($n_{\rm S}$) needed to detect an extended source by computing the full Poisson probability, $P_{\rm null}=1-\sum\limits_{m=0}^{n_{\rm T}-1} \frac{n_{\rm B}^m}{m!} {\rm e}^{-n_{\rm B}}$, with a detection limit of $P_{\rm null}<2.56 \times 10^{-12}$. $n_{\rm B}$ and $n_{\rm T}$ denote the number of background and total detected counts, respectively. We use the same background count rate as that of \citet{2012arXiv1209.3114M}, i.e., $7.70 \, \rm cts \, s^{-1} \, deg^{-2} \ in \ 0.5-2 \, keV$, which is consistent with the mean background count rate obtained for the ten pulsars from the ROSAT All-Sky Survey diffuse background maps\footnote{\url{https://heasarc.gsfc.nasa.gov/cgi-bin/Tools/xraybg/xraybg.pl}}. A higher background count rate would certainly result in a higher $n_{\rm S}$, so we need more exposure time to reach the same sensitivity. Once we get the net count rate, a conversion factor $1.4 \times 10^{-12} \rm \, erg \, cm^{-2} \, s^{-1}$ per count rate is multiplied to obtain the final flux limits. In Figure~\ref{fig:1}, the magenta lines represent the corresponding sensitivities for three different exposure times, namely: the first all-sky survey (eRASS:1, 250 s, solid line), four year all-sky survey (eRASS:8, 2 ks, dashed line) and 20 ks pointed observation (dotted line). We find that almost all of the listed pulsar halo candidates could be detected by the eROSITA telescope for eRASS:8. Three of them (PSR J2032+4127, PSR J1928+1746 and PSR J1831$-$0952) could even be detected for eRASS:1. It implies that the eROSITA telescope has the ability of discovering a considerable number of X-ray halos around middle-aged pulsars.

The electron spectrum in the pulsar halo is not well known. A hard spectrum $p = 1.6$ is favored in the work of \citet{2019ApJ...878..104X} in order to satisfy the GeV upper limit of Geminga from the analysis of {\it Fermi}-LAT data. This value is significantly smaller than $p = 2.34$ used in other works \citep[e.g.,][]{2017Sci...358..911A, 2019MNRAS.484.3491T, 2019PhRvD.100l3015D}. Here we demonstrate that the X-ray halo flux is insensitive to the value of $p$ when the TeV flux is fixed. As shown in Figure~\ref{fig:2}, we plot the IC and synchrotron emission for three different spectral indexes ($p=$1.6, 1.9, and 2.2) for Geminga while fitting the HAWC data. It is clear that with increasing $p$, the GeV flux gets higher. However, in the eROSITA energy range, the flux is nearly independent of $p$. This is  because the electrons emitting X-ray photons in $\rm 0.5 - 2 \, keV$ are just the same ones that  produce $\gamma$ -ray photons in $\rm 8 - 40 \, TeV$ through the IC process.

\subsection{Spatial extension}

The size of X-ray synchrotron emission around pulsar is crucial for identifying whether it is an X-ray halo or not. To make an estimate of the radial extension, we take Geminga and PSR J1928+1746 for example, and calculate their X-ray flux and intensity profile in $\rm 0.5 - 2 \, keV$ as a function of radius, which are shown in Figure~\ref{fig:3}, where we also plot the eROSITA sensitivity curves. The intensity sensitivity is calculated from each concentric annulus in the celestial sphere, which is dependent on the choice of $\Delta R = R_{\rm outer} - R_{\rm inner}$. We take $\Delta R$ to be 5 times that of { the spatial resolution of the telescope} (i.e., $\Delta R = 1.5^{\prime}$) for instance when calculating the sensitivity of eROSITA shown in Figure~\ref{fig:3}. Using a larger $\Delta R$ would increase the detectability of each annulus, but too large a $\Delta R$ would lose the information of the profile. As shown in the upper panel, the intensity profile of the X-ray halos could be measured by eROSITA to a radius larger than 10 pc for Geminga and to a radius larger than 30 pc for PSR J1928+1746 given such a choice of $\Delta R$.  The figure in the lower panel shows that as a whole, eROSITA could detect X-ray halos from both Geminga and PSR 1928+1746, with a radius out to 30\,pc. 

\subsection{Constraining the properties of pulsar halos}

In the previous sections, we calculated the SEDs and spatial extensions for the pulsar halo candidates listed in table \ref{tab:1}. Here, we study the influence of some model parameters on the results obtained above and how the eROSITA telescope could constrain their values. We take Geminga as a reference case since it is one of the well-studied pulsar halos. The results can be similarly applied to other pulsar halos.

\subsubsection{Magnetic field}

In the above, we have assumed  that the value of magnetic field in the pulsar halo is the same as that in the interstellar medium (i.e. $B = \rm 3 \, \mu G$). However, the magnetic field in the pulsar halos is largely unknown. \citet{2019ApJ...875..149L} found the magnetic field inside the TeV halo of Geminga is required to be $<1\, \mu$G in order to satisfy the X-ray flux upper limit derived from the observations of XMM-Newton and Chandra. Since the power of synchrotron radiation is proportional to the energy density of magnetic field, varying magnetic field could dramatically affect the X-ray flux. 

To investigate how deep eROSITA can constrain the magnetic field in the pulsar halos, we calculate the X-ray fluxes for different magnetic field as a function of angular radius while simultaneously fitting the spectrum and spatial profile of the TeV emission measured by HAWC, which are shown in Figure~\ref{fig:4}. In Figure~\ref{fig:4}a, we gradually increase the magnetic field until the flux curve just touch the eROSITA sensitivity curves, which is the critical value of the magnetic field below which eROSITA is unable to see the X-ray halo under the correspondent exposure time. Figure~\ref{fig:4}a shows that eROSITA is able to constrain the magnetic filed down to $\rm 0.6 \, \mu G$ for an exposure time of 20 ks if the X-ray halo is not detected. 

\subsubsection{Diffusion coefficient}

The value of diffusion coefficient affects the spatial distribution of electrons and hence the intensity profile of pulsar halos. In section \ref{subsec:SEDs}, we used a low diffusion coefficient ($D_0 = 3.5 \times 10^{27}  \rm \, cm^2 \, s^{-1}$), which is about two orders of magnitude smaller than the typical value in the ISM. The origin of such low diffusion zone around Geminga is still unclear, though various theoretical explanations have been proposed, e.g., a small turbulence injection scale \citep{2018MNRAS.479.4526L}, self-confinement of CRs \citep{2018PhRvD..98f3017E}, SNR environment \citep{2019MNRAS.488.4074F} and  anisotropic diffusion\citep{2019PhRvL.123v1103L}.

As shown in Figure~\ref{fig:4}c, a lower diffusion coefficient is required for a lower magnetic field to steepen the TeV intensity profile. This is because a lower magnetic field makes the TeV intensity profile flat in the inner region due to a longer cooling timescale of electrons (see \citealt{2019ApJ...875..149L} for a quantitative discussion about this). Therefore, with TeV observations alone, it is unable to determine the magnetic field and diffusion coefficient separately. However, different combinations of $B$ and $D_0$ could significantly affect the X-ray intensity profile (see Figure~\ref{fig:4}d), thus their values are expected to be well constrained with future observations of the eROSITA telescope. We note that the diffusion coefficient we used is smaller than the Bohm diffusion coefficient for $B = 1.2 \, \rm \mu G \ and \ 0.6 \, \mu G$, which could indicate that the assumption of  homogeneous diffusion in the whole pulsar halo may not be realistic. Assuming an inhomogeneous diffusion could solve this problem \citep{2019ApJ...875..149L}, but  a detailed discussion on this issue is beyond the scope of the present work. 

The diffusion coefficient is presumed to have an energy dependence in the form of power law function with a spectral index $\delta = 1/3$ (i.e. $D=D_0 (E/100 \rm \, TeV)^{1/3}$), which is motivated by the Kolmogorov turbulence model \citep{1941DoSSR..30..301K}. However, other models suggest a different $\delta$, e.g., $\delta = 1/2$ by \citet{1965PhFl....8.1385K}. Here, we consider four cases: $\delta = 0, \, 1/3, \, 1/2$ and 1, with other parameters the same as that in Figure~\ref{fig:1}, and calculate their corresponding intensity profiles in $\rm 0.5-10 \, keV$ for Geminga. The results are shown in Figure~\ref{fig:5}. We divide the energy range of eROSITA into two X-ray bands: soft band ($\rm 0.5-2  \, keV$) and hard band ($\rm 2-10 \, keV$). We find that in the soft X-ray band the intensity profile is insensitive of $\delta$. This is because the soft X-ray photons are radiated by the same electrons that produce the HAWC emission (i.e. $\sim \rm 100 \, TeV$). However, the difference between X-ray intensity profiles becomes pronounced in the hard X-ray band, since the diffusion coefficient of the hard X-ray emitting electrons varies significantly with $\delta$. The difference may be resolved by the eROSITA telescope with its good angular resolution and sensitivity, so eROSITA has the ability to test the existing theories for diffusion coefficient power index $\delta$.

\section{Discussion and Summary} \label{sec:discussion and summary}

In the above calculations we have assumed that  electrons are injected from a stationary point in the space, so the electron distribution is isotropic considering the isotropic diffusion of electrons. However, pulsars may have proper motion which could result in an anisotropic distribution of electrons, with the electron density enhanced along the direction of pulsar motion \citep{2020arXiv201015731Z}. We note that the effect of proper motion is unimportant in this work because the X-ray emitting electrons  have a much shorter lifetime compared to the age of middle-aged pulsars and thus only the recently-injected electrons contribute to the observed X-ray flux, quite similar to the case of TeV gamma-ray emission from pulsar halos \citep{2020arXiv201015731Z}. Besides, the direction of mean magnetic field inside the pulsar halo could lead to an anisotropic diffusion of electrons and thus an anisotropic X-ray profile \citep{2019PhRvL.123v1103L}, which is, however, beyond the scope of this paper. 

In section \ref{subsec:SEDs}, we calculated the SEDs for pulsars with their IC fluxes normalized to the  HAWC data reported in the third HAWC catalog of very-high-energy gamma-ray sources \citep{2020ApJ...905...76A}, where the TeV flux of each source is calculated based on  four different hypothetical morphologies (i.e., point sources, and extended disk-like sources with radii of $0.5^\circ$, $1.0^\circ$, and $2.0^\circ$). The real flux for more extended source could be larger than that given in the HAWC catalog. Therefore, we may underestimate the IC and synchrotron emission for some pulsar halos in table \ref{tab:1}. Note that this does not apply to  Geminga and Monogem halos, where the diffusion model has been employed in the data analysis (HAWC data of Geminga and Monogem are taken from \citet{2017Sci...358..911A}). More detailed morphological studies of TeV halos are needed in the future for a more precise calculation of the X-ray halo flux.

In summary, we studied the possible X-ray halo emission around 10 middle-aged pulsars with the highest $\dot E / d^2$ ratios assuming an interstellar magnetic field of $3\mu\,$G. By comparing the predicted X-ray flux with the estimated sensitivities of the eROSITA telescope, we found that most of these pulsar halos could be detected by eROSITA with its four year all-sky survey. Three of them may have been already detectable with the first-year all-sky survey data of eROSITA. The intensity profile of the X-ray halo emission from some pulsars (e.g. Geminga and PSR J1928+1746) could be detected up to a radius of several tens of pc by eROSITA. Future observations of the intensity profiles of X-ray halos with the eROSITA telescope can give stringent constraints on the magnetic field and the properties of the diffusion coefficient in the pulsar halos.

\section*{Acknowledgements}
This work is supported by the National Key R \& D program of China under the grant 2018YFA0404203, NSFC grants 11625312, 11851304,
and U2031105.

\section*{Data Availability}
The data underlying this article will be shared on reasonable request to the corresponding author.

\bibliographystyle{mnras}
\bibliography{X_ray_halo}

\newpage
\begin{table*}
	\centering
	\caption{X-ray halo candidate pulsars. Pulsar properties are taken from the ATNF pulsar catalogue \citep{2005AJ....129.1993M}. }
	\label{tab:1}
	\begin{tabular}{lccccccl}
		\hline
		Pulsar name & R.A. & Decl. & d & $\tau_c$ & $\dot E_{\rm spin}$ & $\dot E_{\rm spin}/d^2$ & Nearest 3HWC source\\
		   & (deg) & (deg) & (kpc) & (kyr) & ${\rm (10^{34} \, erg \, s^{-1})}$ & ${\rm (10^{34} \, erg \, cm^{-2} \, s^{-1})}$ &  \\
		\hline
		J0633+1746 & 98.47  & 17.77 & 0.25 & 342  & 3.2  & 51.2  & 3HWC J0634+180$^{(a)}$\\
		B0656+14   & 104.95 & 14.24 & 0.29 & 111  & 3.8  & 45.18 & 3HWC J0702+147\\
		B1951+32   & 298.24 & 32.88 & 3    & 107  & 374  & 41.56 & 3HWC J1951+293\\
		J1954+2836 & 298.58 & 28.6  & 1.96 & 69.4 & 105  & 27.33 & 3HWC J1954+286\\
		J1740+1000 & 265.11 & 10    & 1.23 & 114  & 23.2 & 15.33 & 3HWC J1739+099\\
		J1913+1011 & 288.33 & 10.19 & 4.61 & 169  & 287  & 13.5  & 3HWC J1912+103\\
		J2032+4127 & 308.05 & 41.46 & 1.33 & 201  & 15.2 & 8.59  & 3HWC J2031+415\\
		J1928+1746 & 292.18 & 17.77 & 4.34 & 82.6 & 160  & 8.49  & 3HWC J1928+178\\
		J1831-0952 & 277.89 & -9.87 & 3.68 & 128  & 108  & 7.97  & 3HWC J1831$-$095\\
		J0633+0632 & 98.43  & 6.54  & 1.35 & 59.2 & 12   & 6.58  & 3HWC J0634+067\\
		\hline
	\end{tabular}
	\\ \flushleft{$^{(a)}$It is believed that five sources (3HWC J0630+186, 3HWC J0631+169, 3HWC J0633+191, 3HWC J0634+165, and 3HWC J0634+180) are associated with J0633+1746 (Geminga) \citep{2020ApJ...905...76A}.}
\end{table*}

\newpage
\begin{figure*}
	\centering
	\includegraphics[scale=0.52]{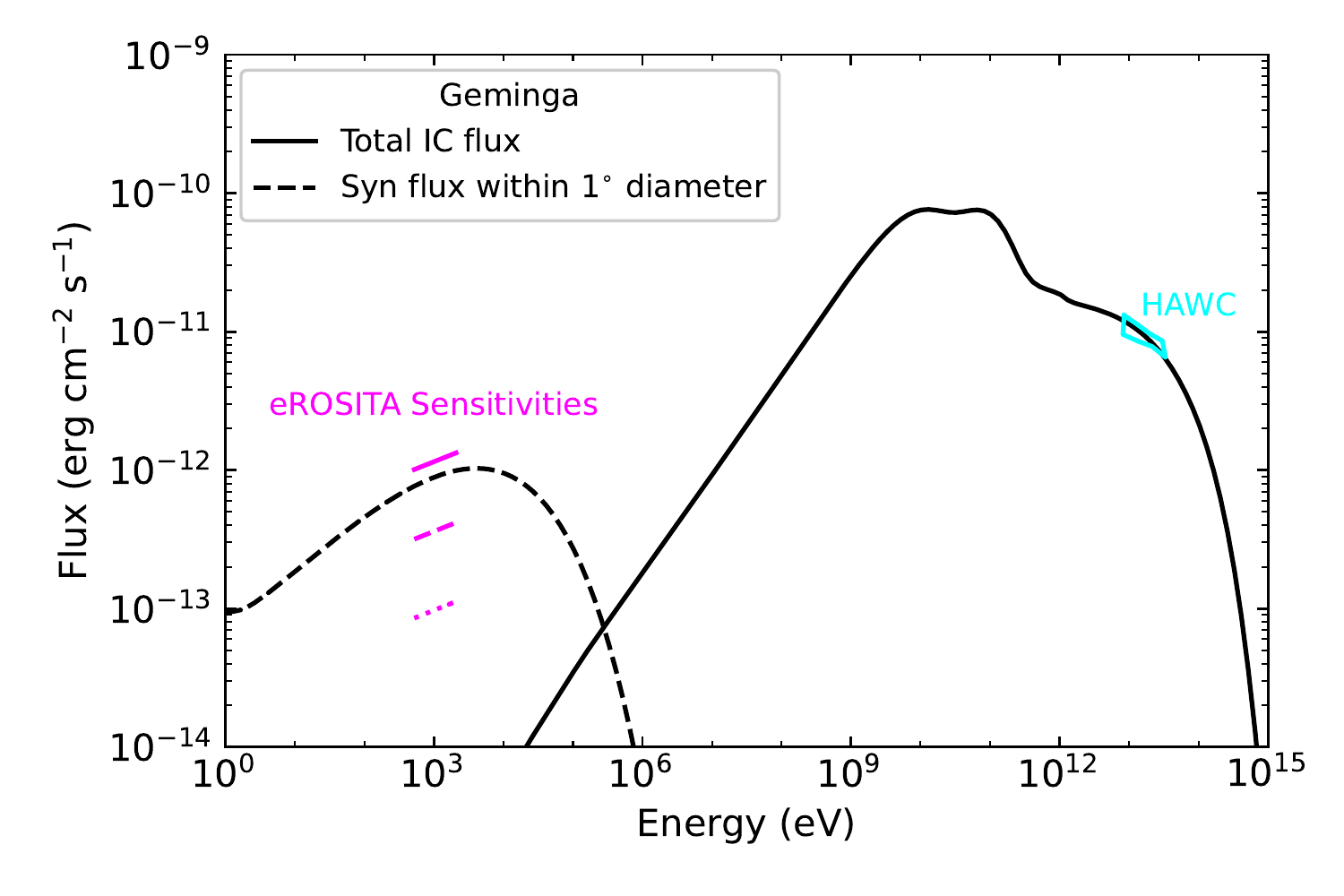}
	\includegraphics[scale=0.52]{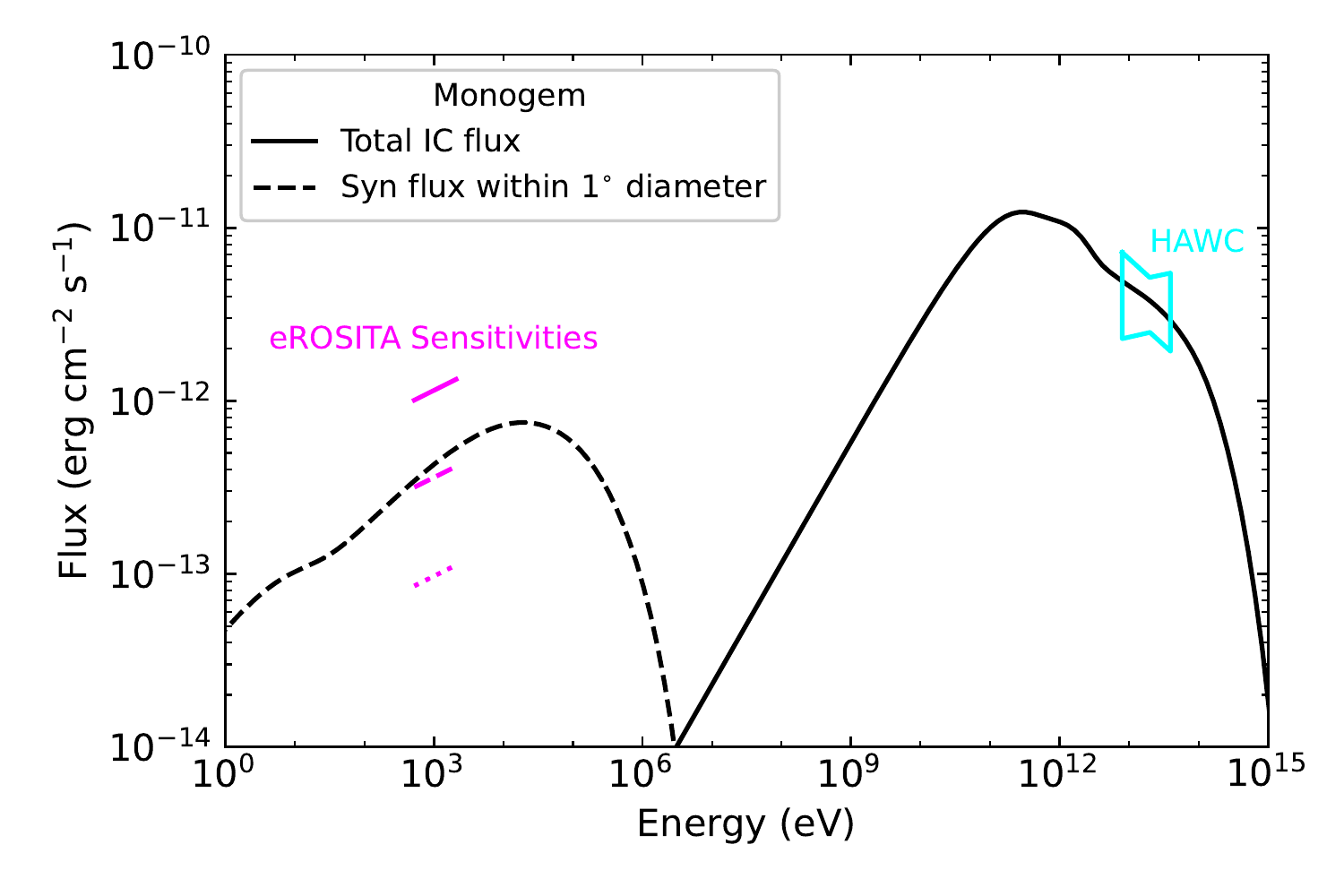}
	\includegraphics[scale=0.52]{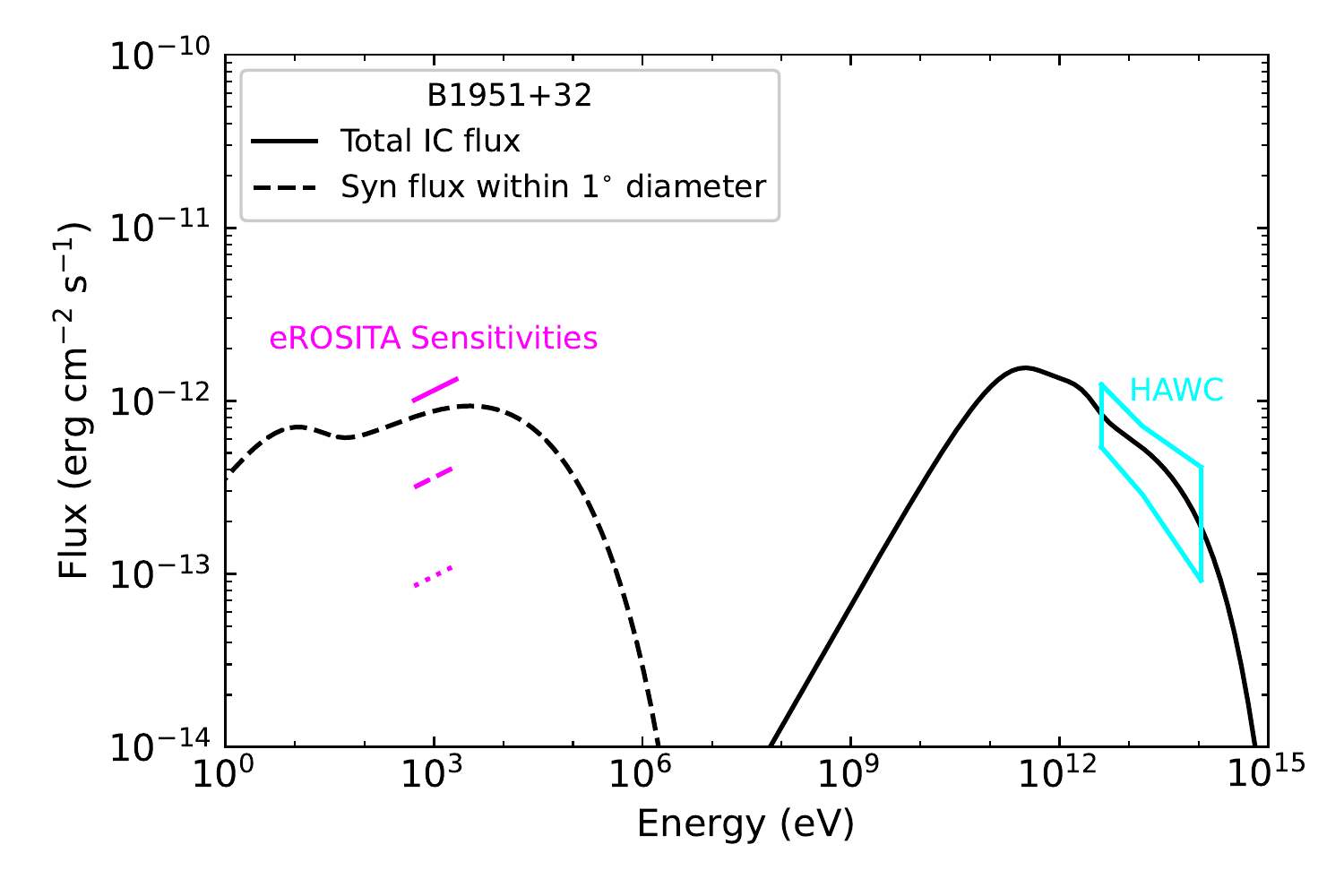}
	\includegraphics[scale=0.52]{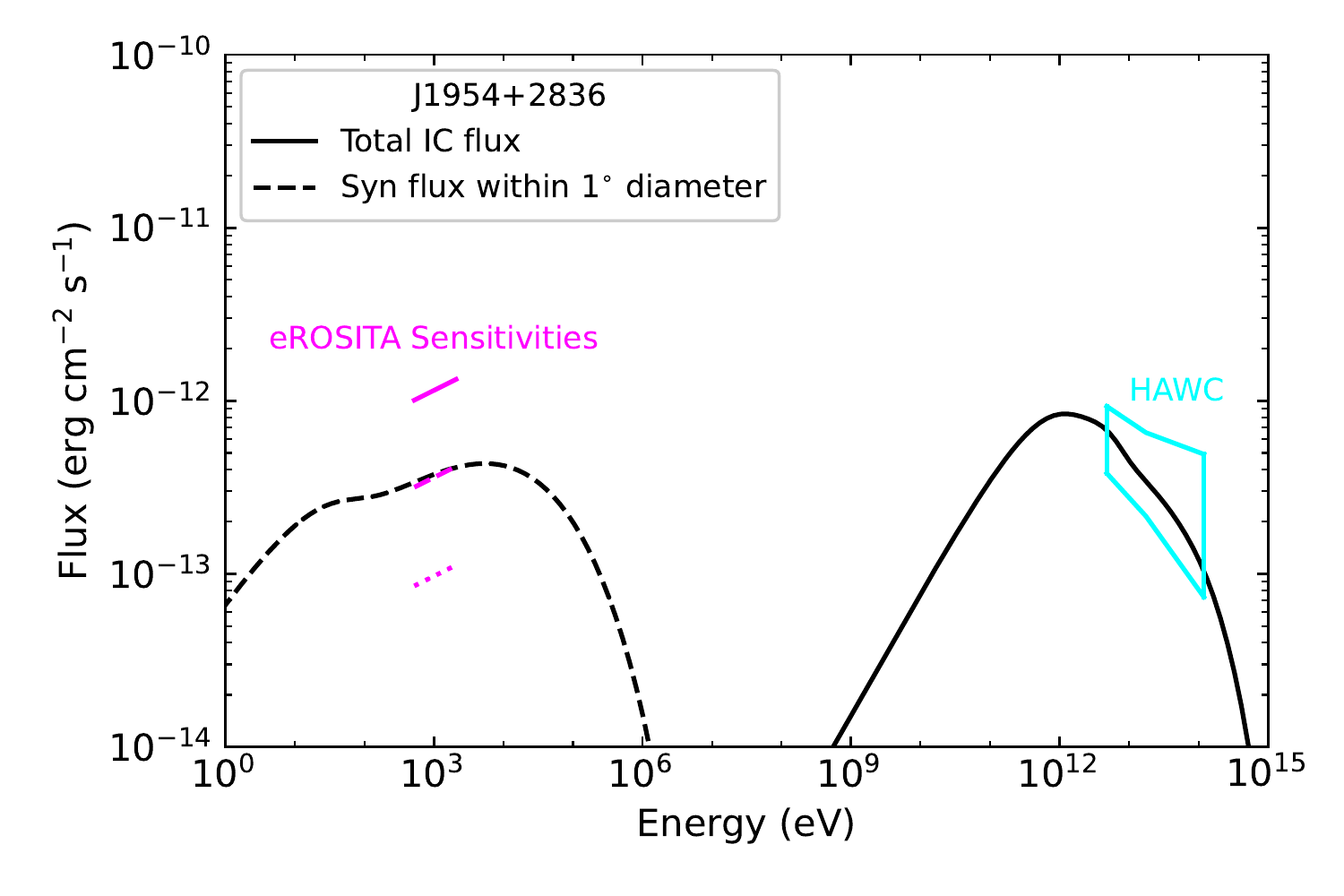}
	\includegraphics[scale=0.52]{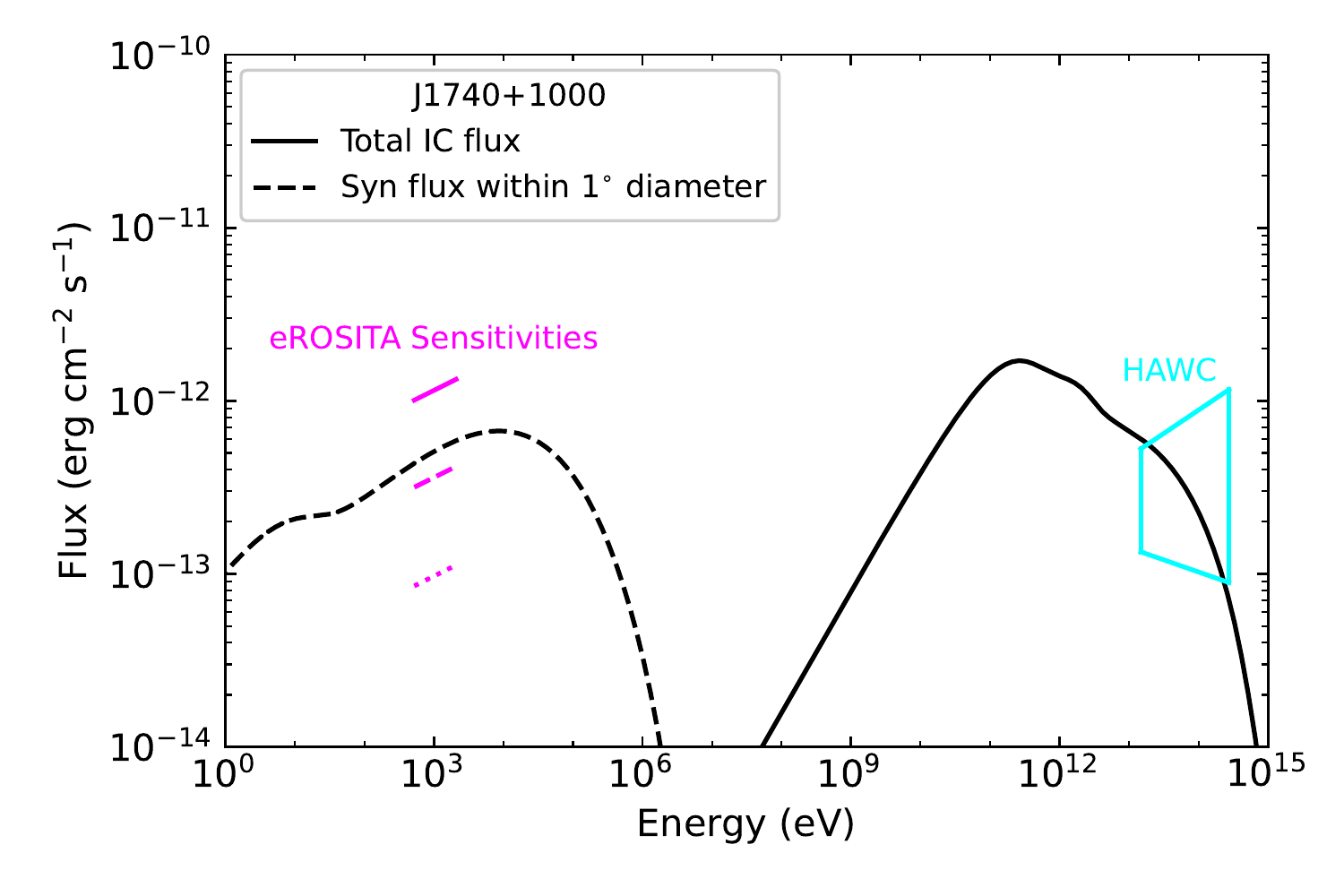}
	\includegraphics[scale=0.52]{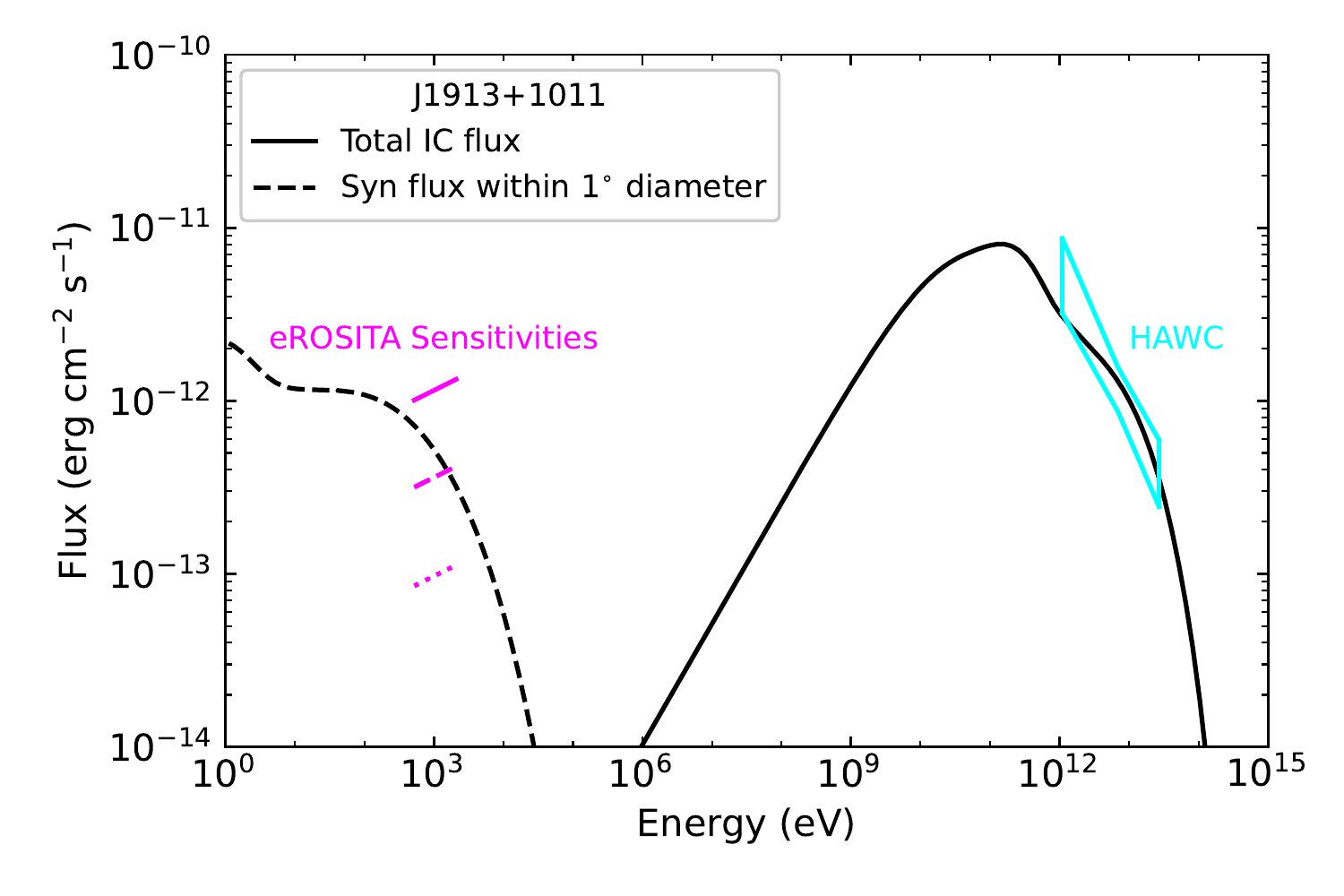}
	\includegraphics[scale=0.52]{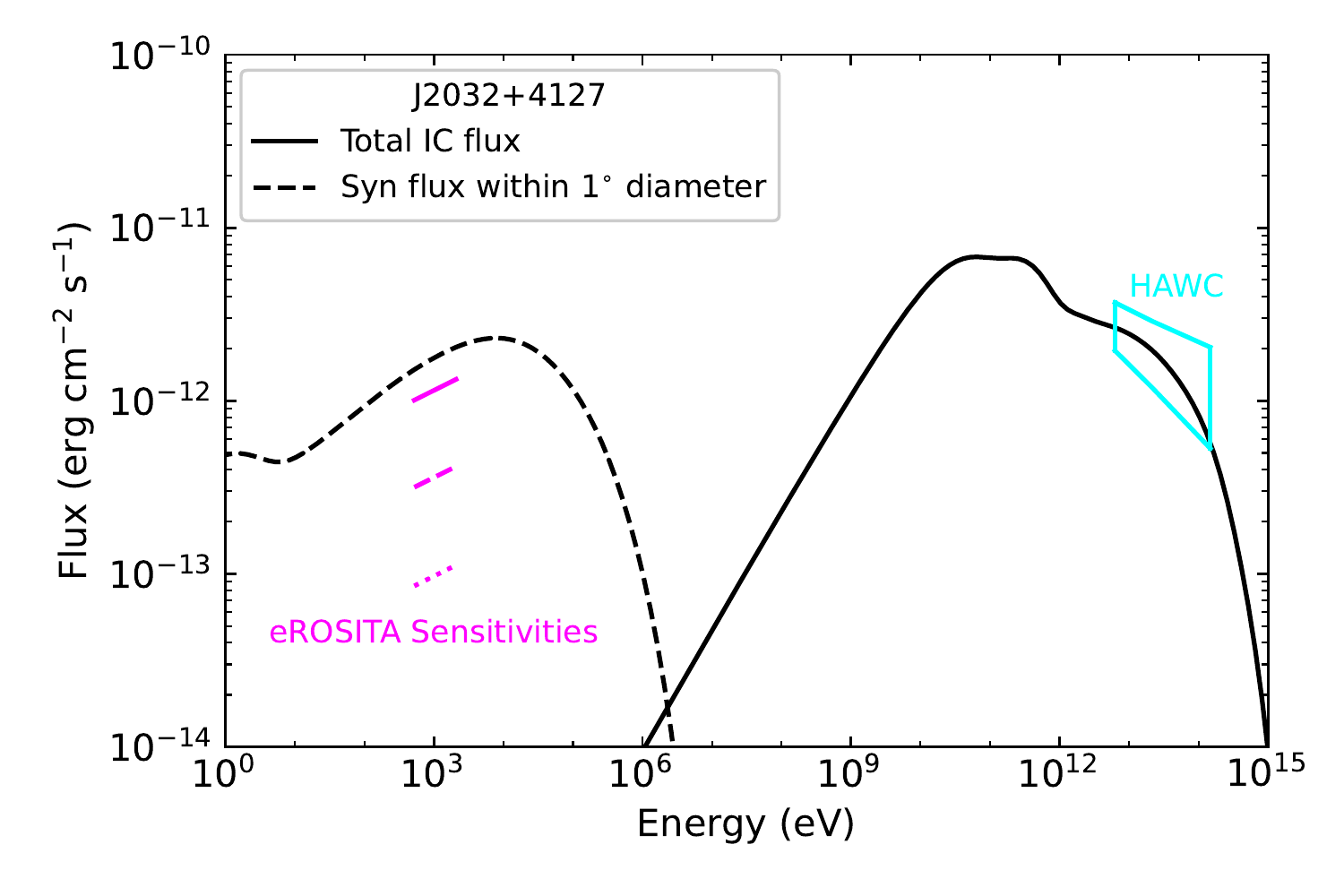}
	\includegraphics[scale=0.52]{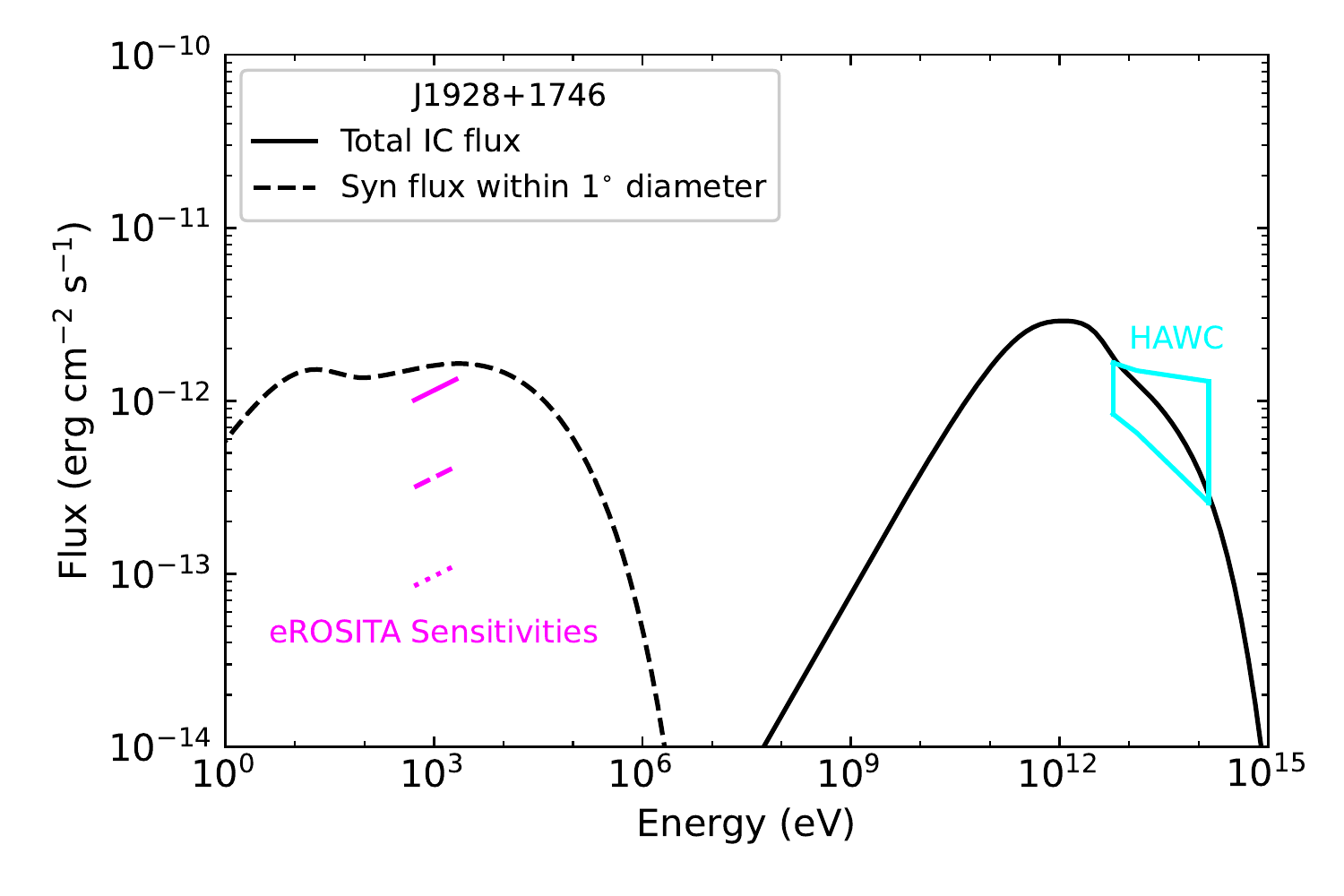}
	\caption{Predicted SEDs of ten pulsar halos listed in table \ref{tab:1}. Model parameters are: $p=1.6$, $B = \rm 3 \, \mu G$, $D_0 = 3.5 \times 10^{27}  \rm \, cm^2 \, s^{-1}$, $\tau_0=12 \, \rm kyr$ and $r_{\rm max}=60 \, \rm pc$. The black solid line is the IC flux within a radius of 60 pc from the pulsar, which is fitted to HAWC data (cyan bow tie). The HAWC data is taken from \citet{2020ApJ...905...76A}, except for Geminga and Monogem, for which we take data from \citet{2017Sci...358..911A}). The black dashed line is the synchrotron flux within an angular radius of $30^{\prime}$ from the pulsar. The  eROSITA sensitivities corresponding to the first all-sky survey (eRASS:1, 250 s), four year all-sky survey (eRASS:8, 2 ks) and 20 ks pointed observation are, respectively, represented by the magenta solid line, dashed line and dotted line.}
	\label{fig:1}
\end{figure*}

\addtocounter{figure}{-1}

\begin{figure*}
	\centering
	\includegraphics[scale=0.52]{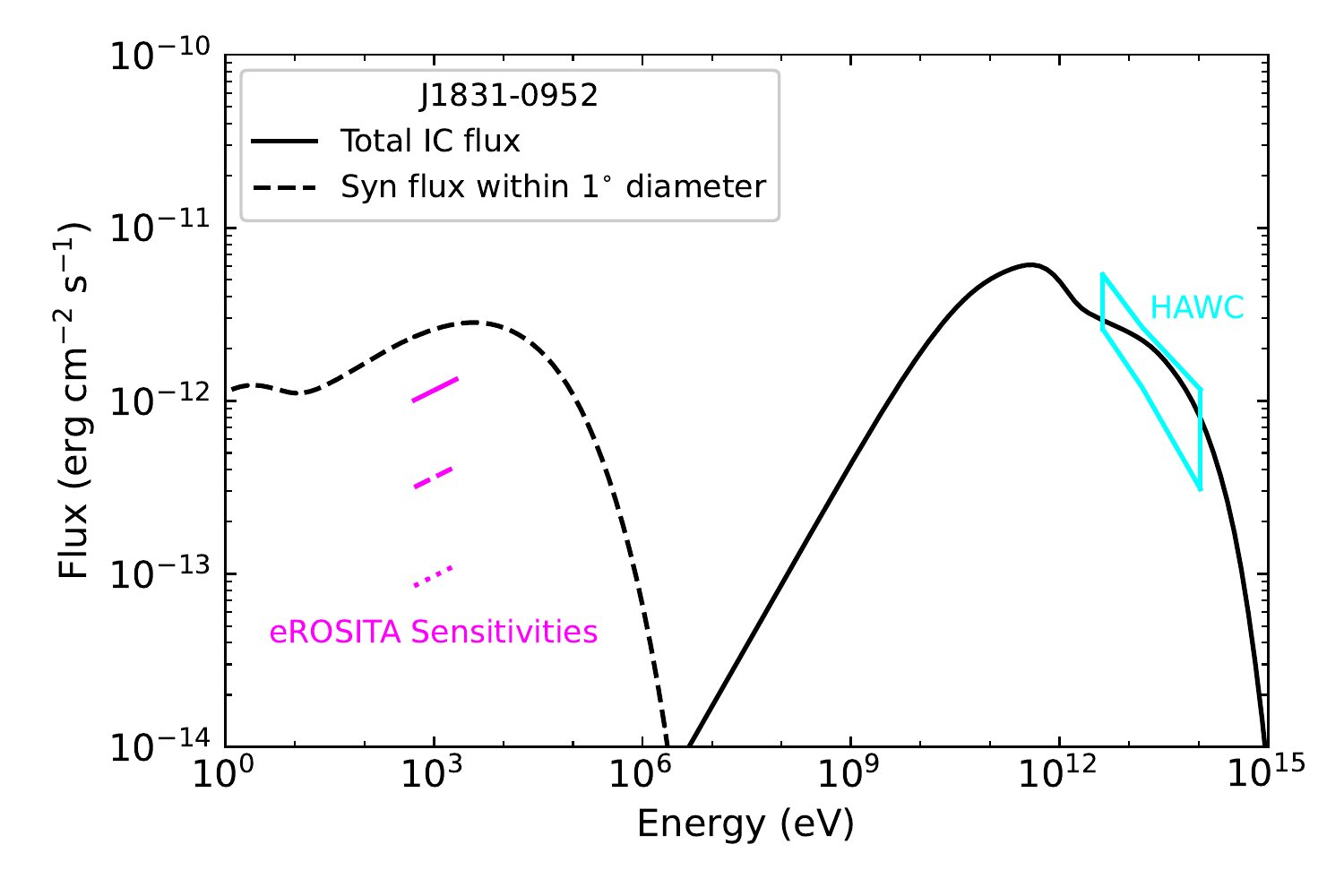}
	\includegraphics[scale=0.52]{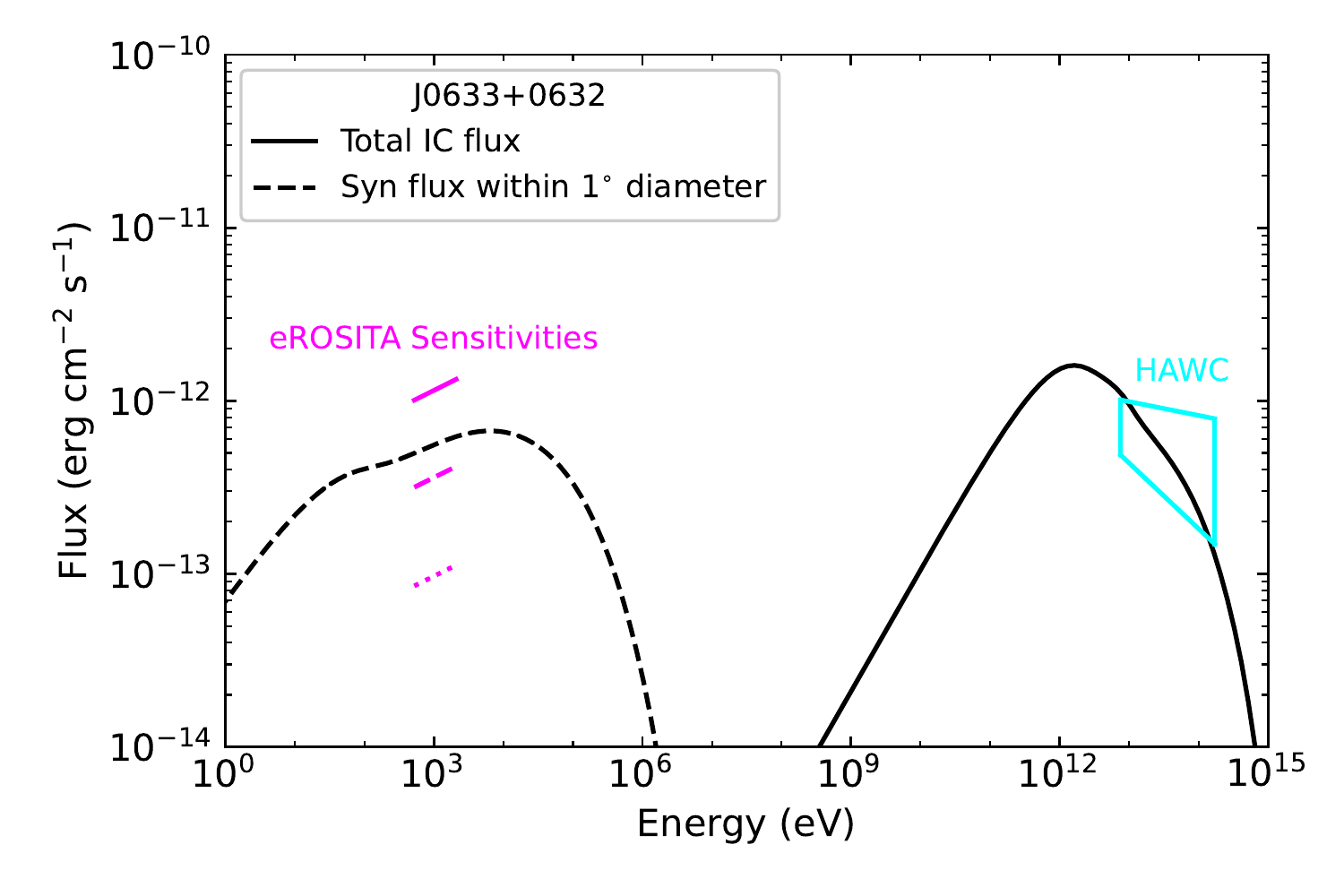}
	\caption{Continued}
\end{figure*}

\begin{figure}
	\centering
	\includegraphics[width=\columnwidth]{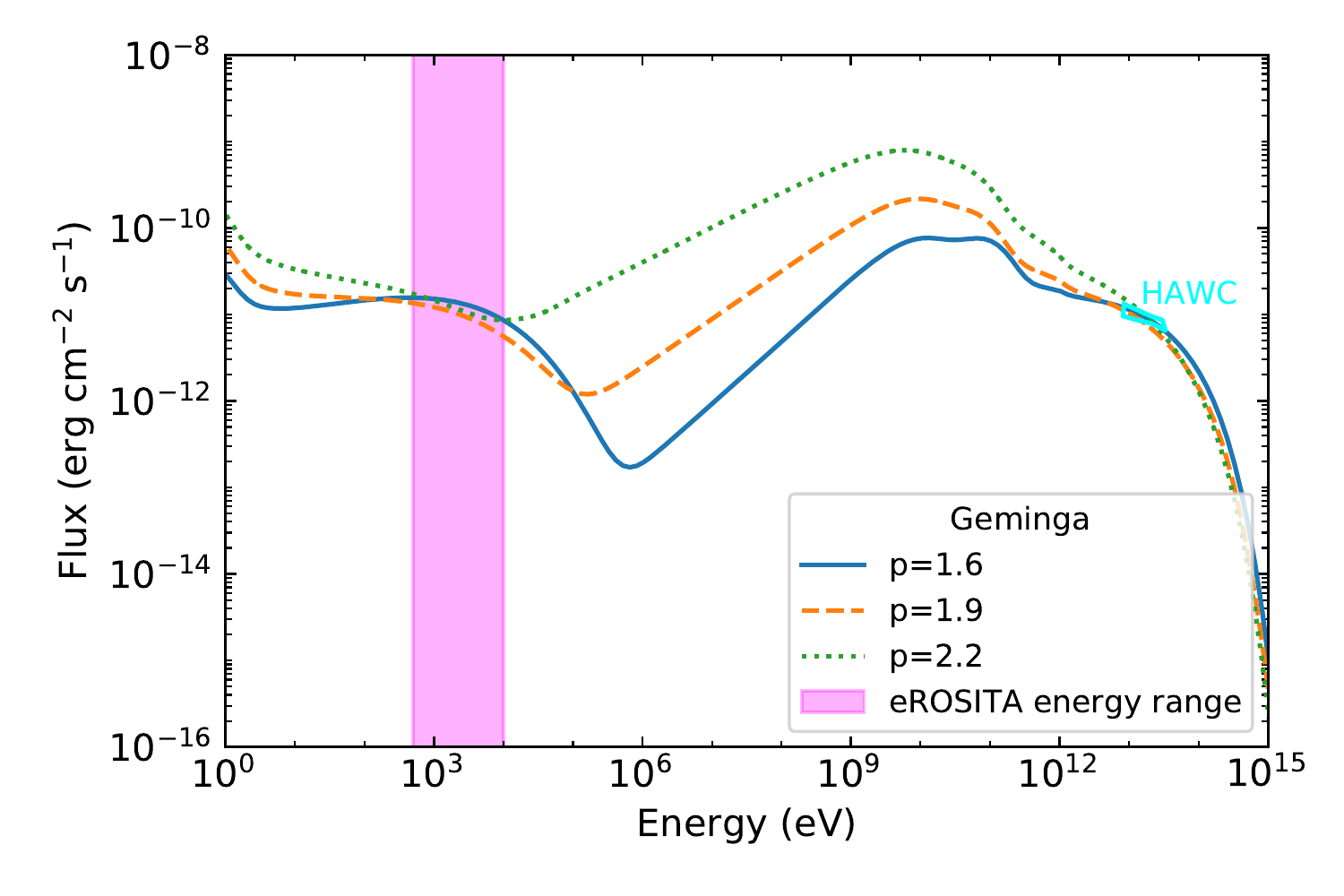}
	\caption{Sum of the synchrotron and IC flux of the halo around Geminga  for three different electron spectral indexes of $p=1.6$ (blue), 1.9 (orange) and 2.2 (green), respectively. The magenta band denotes the eROSITA energy range.}
	\label{fig:2}
\end{figure}

\begin{figure*}
	\centering
	\includegraphics[scale=0.5]{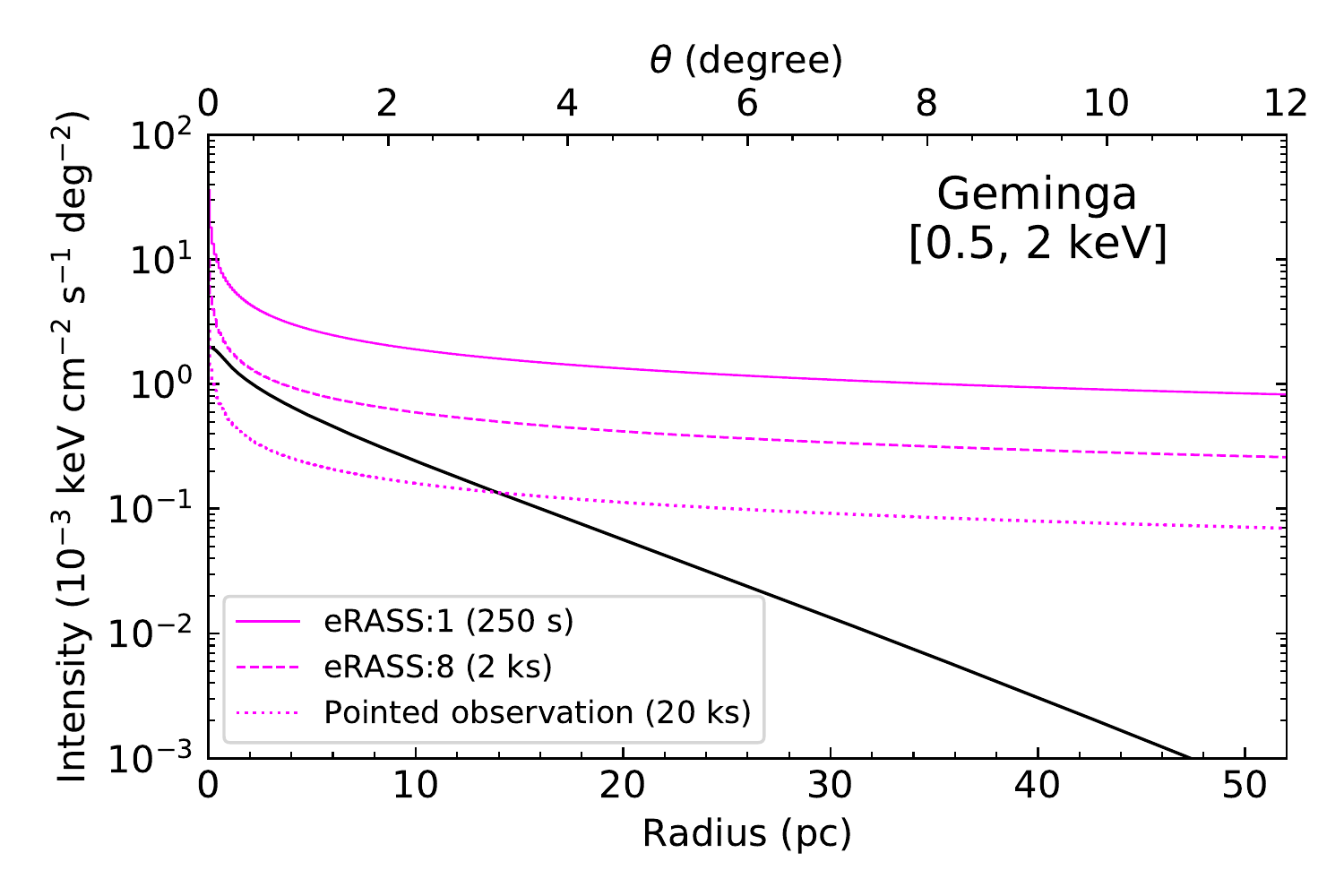}
	\includegraphics[scale=0.5]{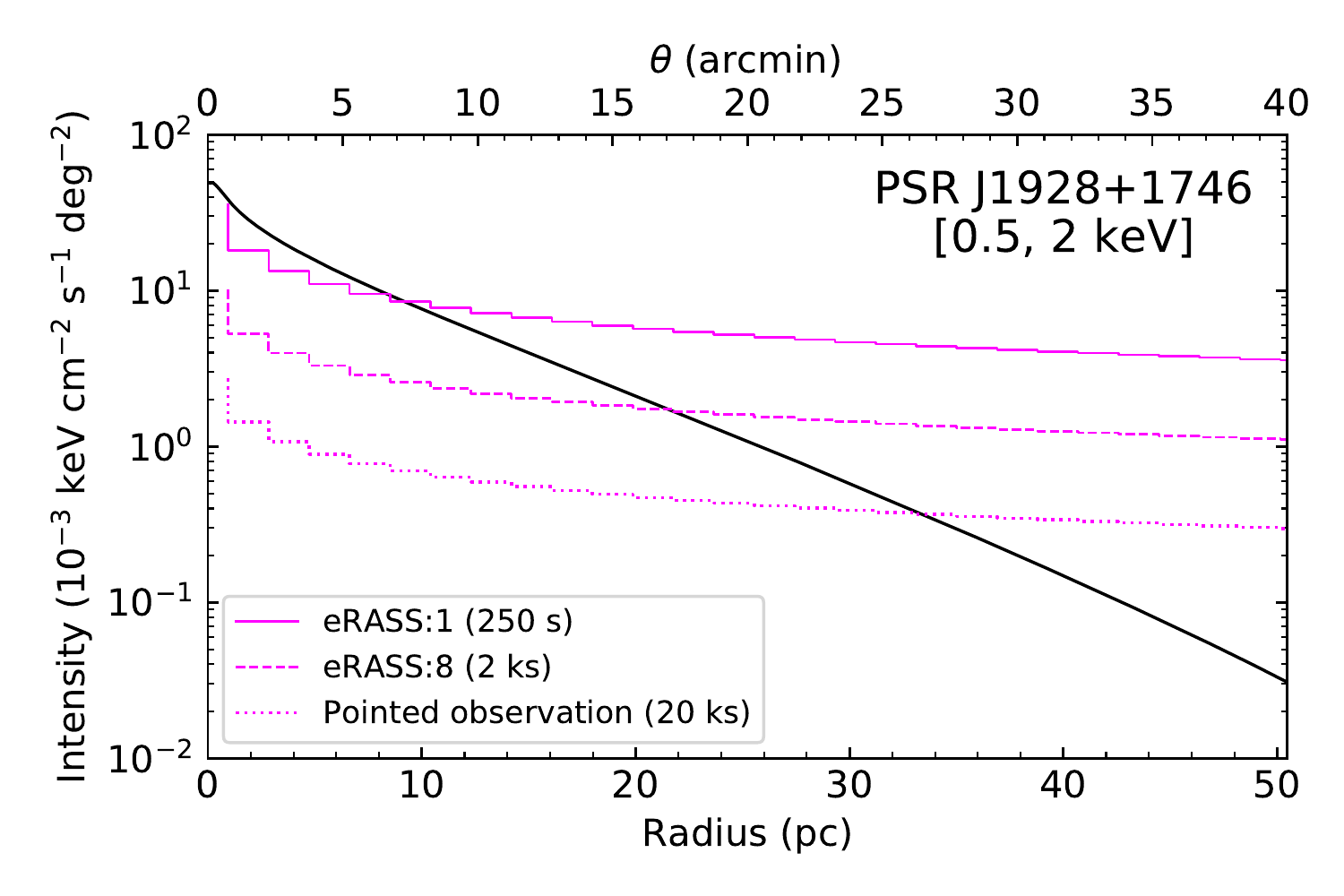}
	\includegraphics[scale=0.5]{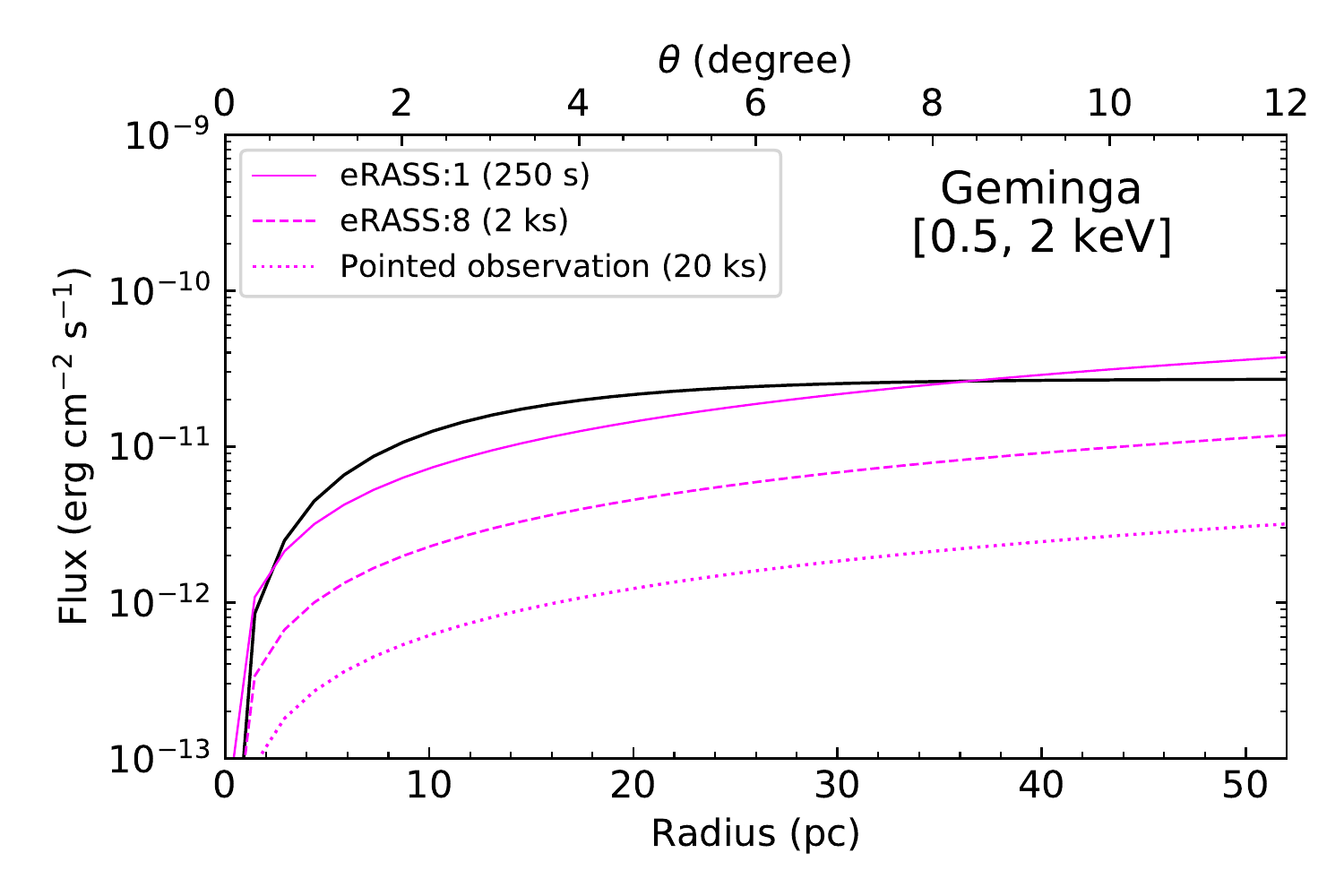}
	\includegraphics[scale=0.5]{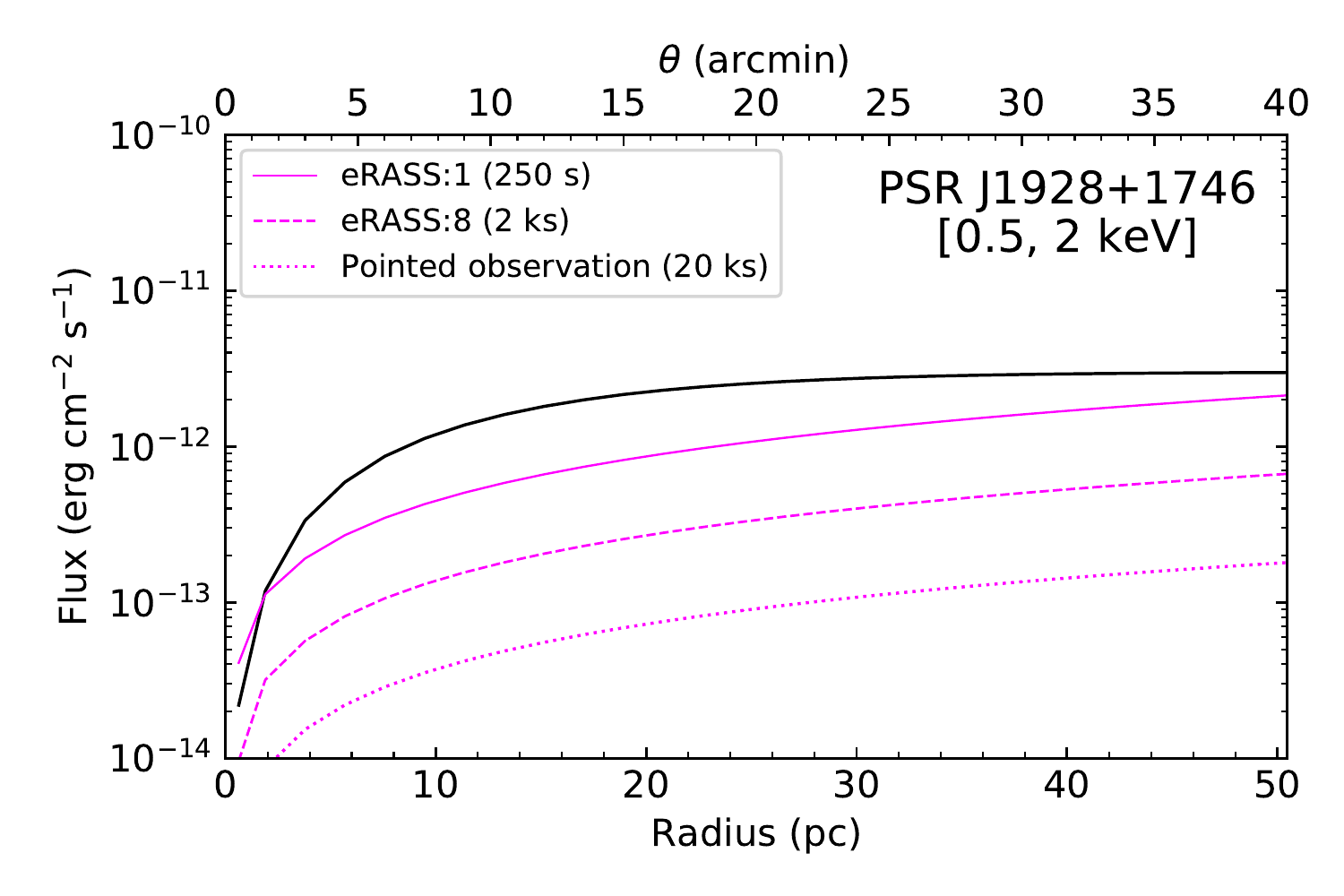}
	\caption{The intensity profile and integrated flux of X-ray halos in $\rm 0.5 - 2 \, keV$ as a function of the distance from Geminga (left) and PSR J1928+1746 (right). Three magenta lines represent the eROSITA sensitivity curves of eRASS:1, eRASS:8 and 20 ks pointed observation, respectively. The eROSITA sensitivity is obtained by taking successive $1.5^{\prime}$ width circular rings as the mapping area, see text for more details.}
	\label{fig:3}
\end{figure*}

\begin{figure*}
	\centering
	\subfigure[]{\includegraphics[scale=0.5]{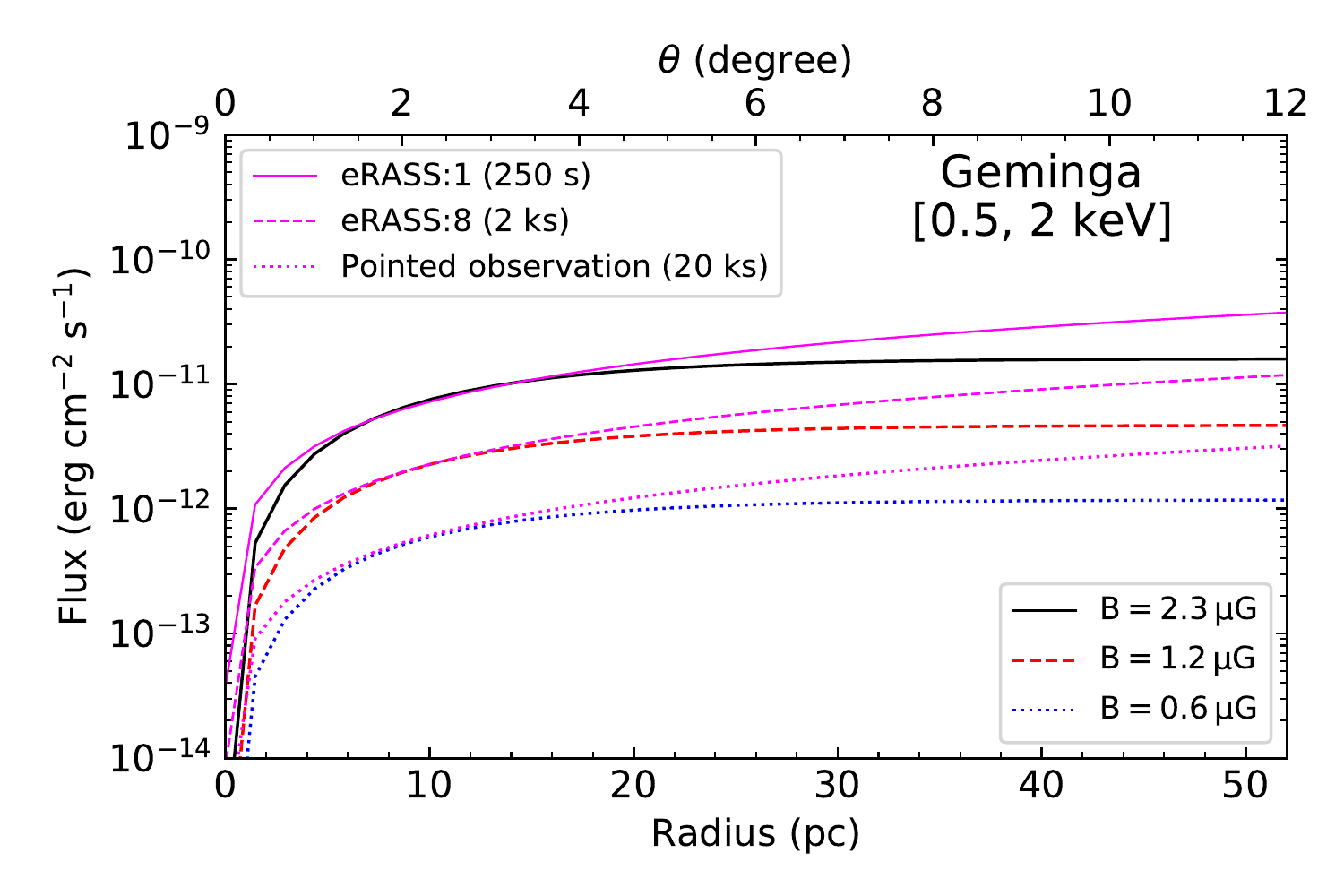}}
	\subfigure[]{\includegraphics[scale=0.5]{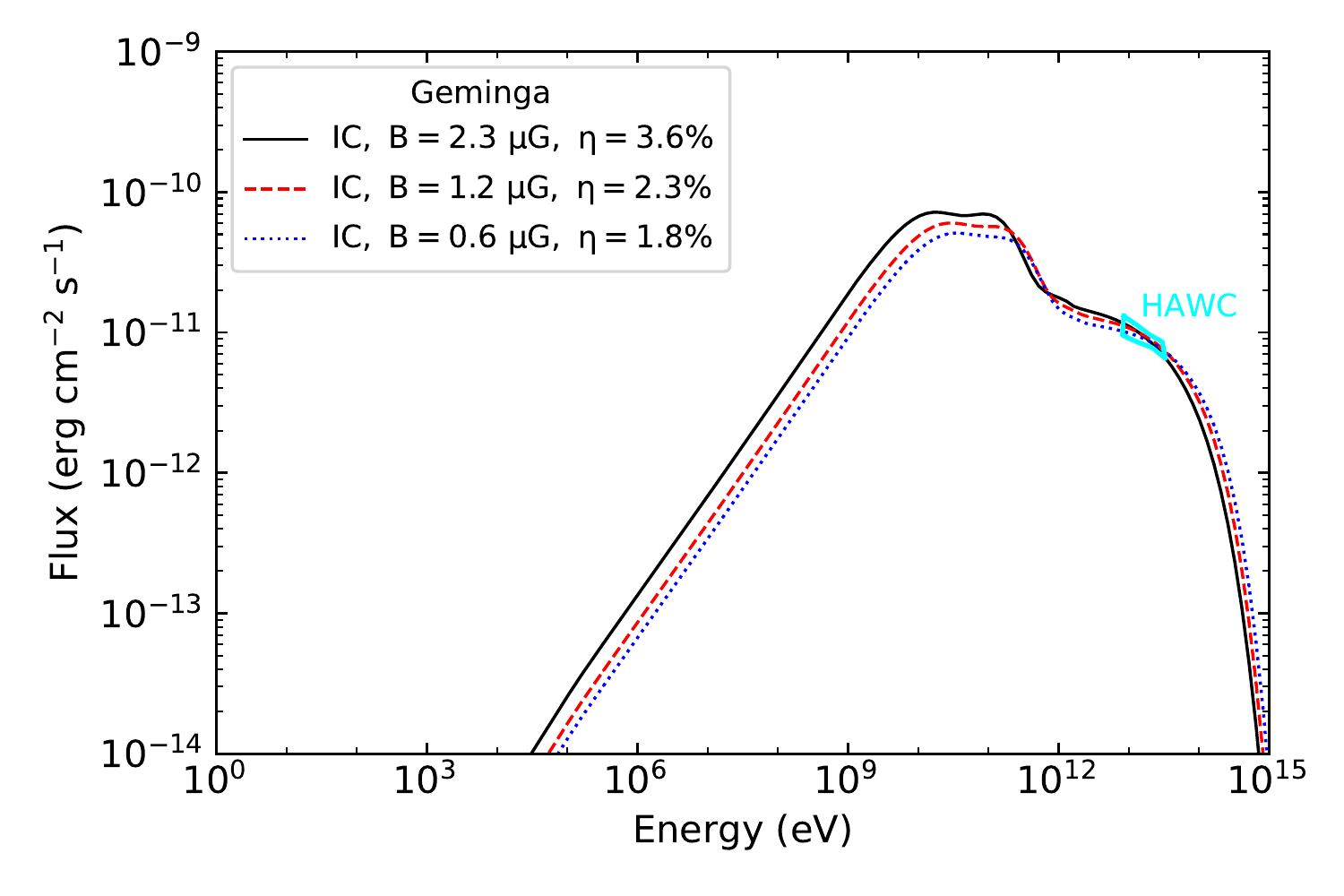}}
		\subfigure[]{\includegraphics[scale=0.5]{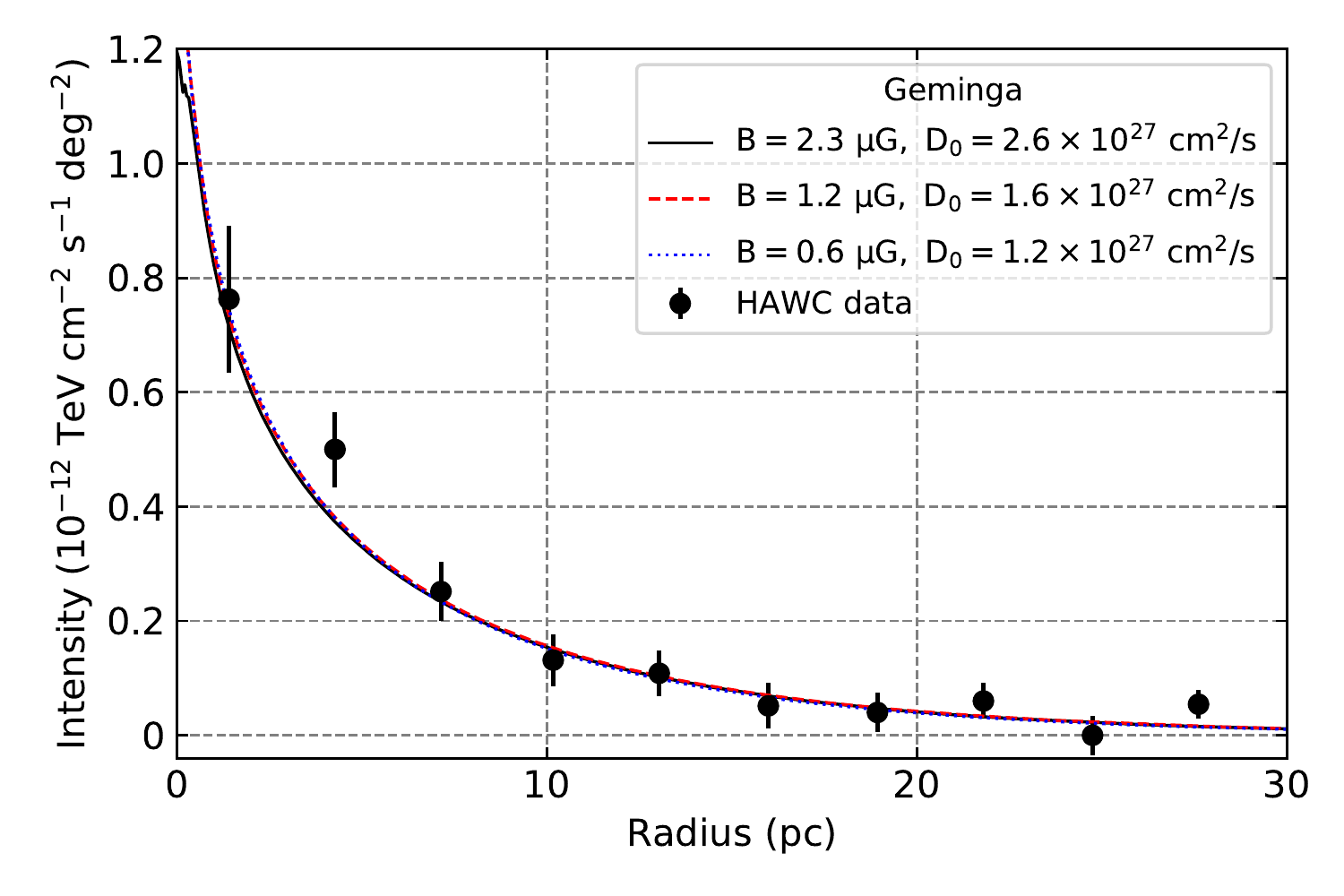}}
		\subfigure[]{\includegraphics[scale=0.5]{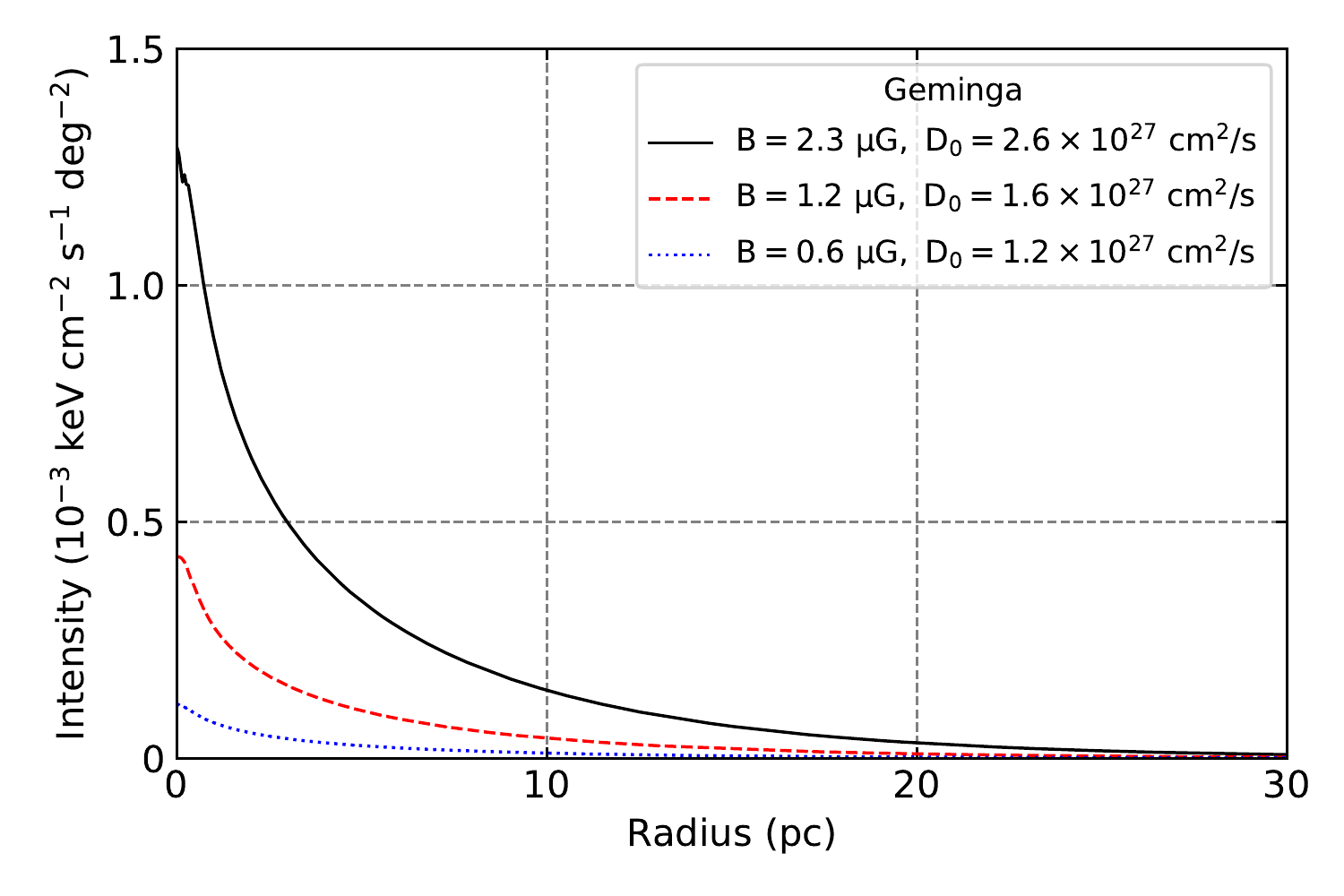}}
	\caption{(a) Integrated flux of the X-ray halo in $\rm 0.5-2 \, keV$ as a function of distance from Geminga. The black solid, red dashed and blue dotted lines represent the predicted fluxes for the magnetic field $B= \rm 2.3 \, \mu G, 1.2 \, \mu G \ and \ 0.6 \, \mu G$, respectively. The magenta lines represent the sensitivity of eROSITA with the same meaning as that in Figure~\ref{fig:3}. (b) IC emission of the pulsar halo of Geminga, fitted to the HAWC flux data in $\rm 8-40 \, TeV$.  (c) Fit of the surface brightness profiles in $\rm 8-40 \, TeV$ measured by HAWC  \citep{2017Sci...358..911A}. (d) Predicted surface brightness profiles in $\rm 0.5-2 \, keV$.}
	\label{fig:4}
\end{figure*}

\begin{figure*}
	\centering
	\includegraphics[scale=0.38]{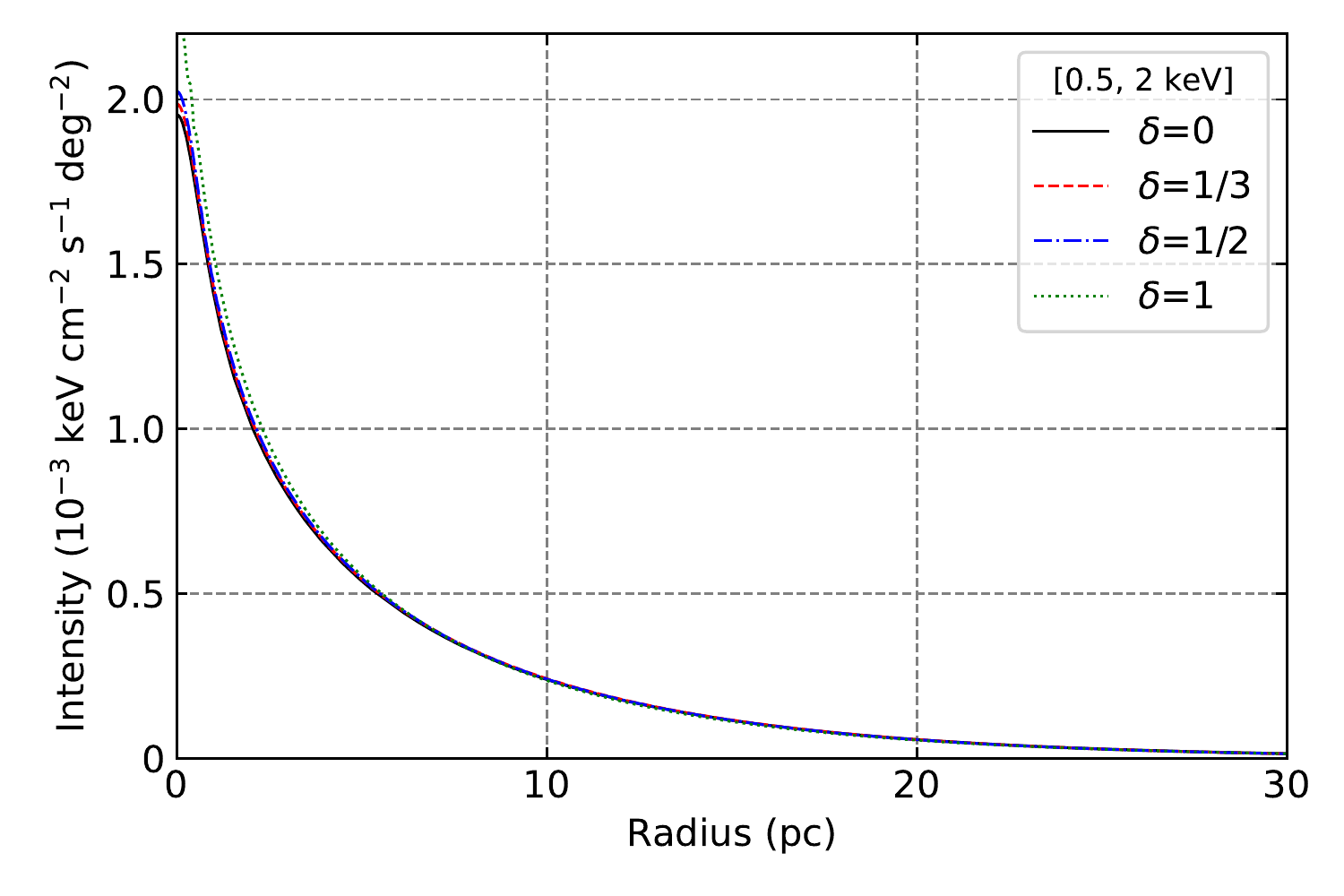}
	\includegraphics[scale=0.38]{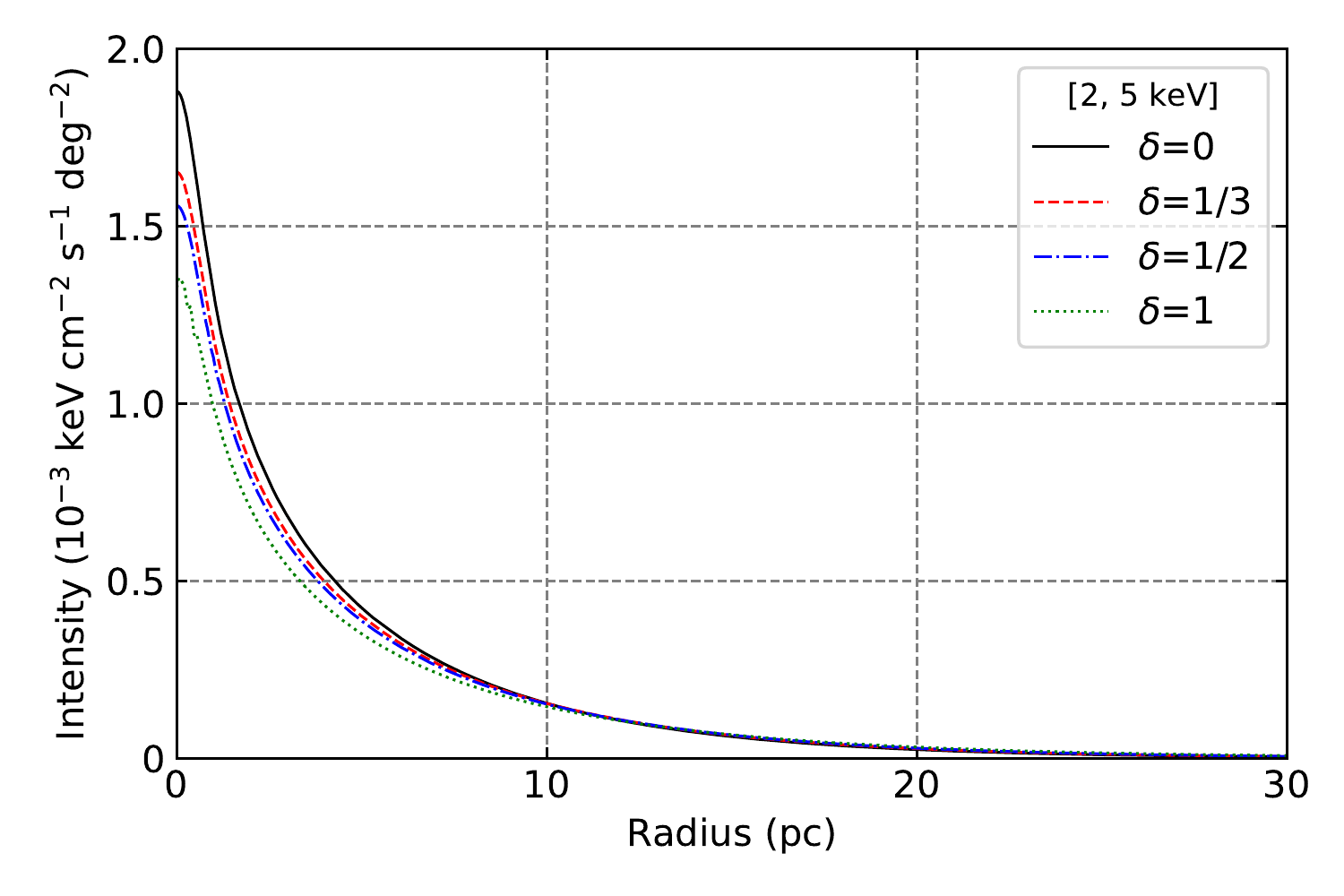}
	\includegraphics[scale=0.38]{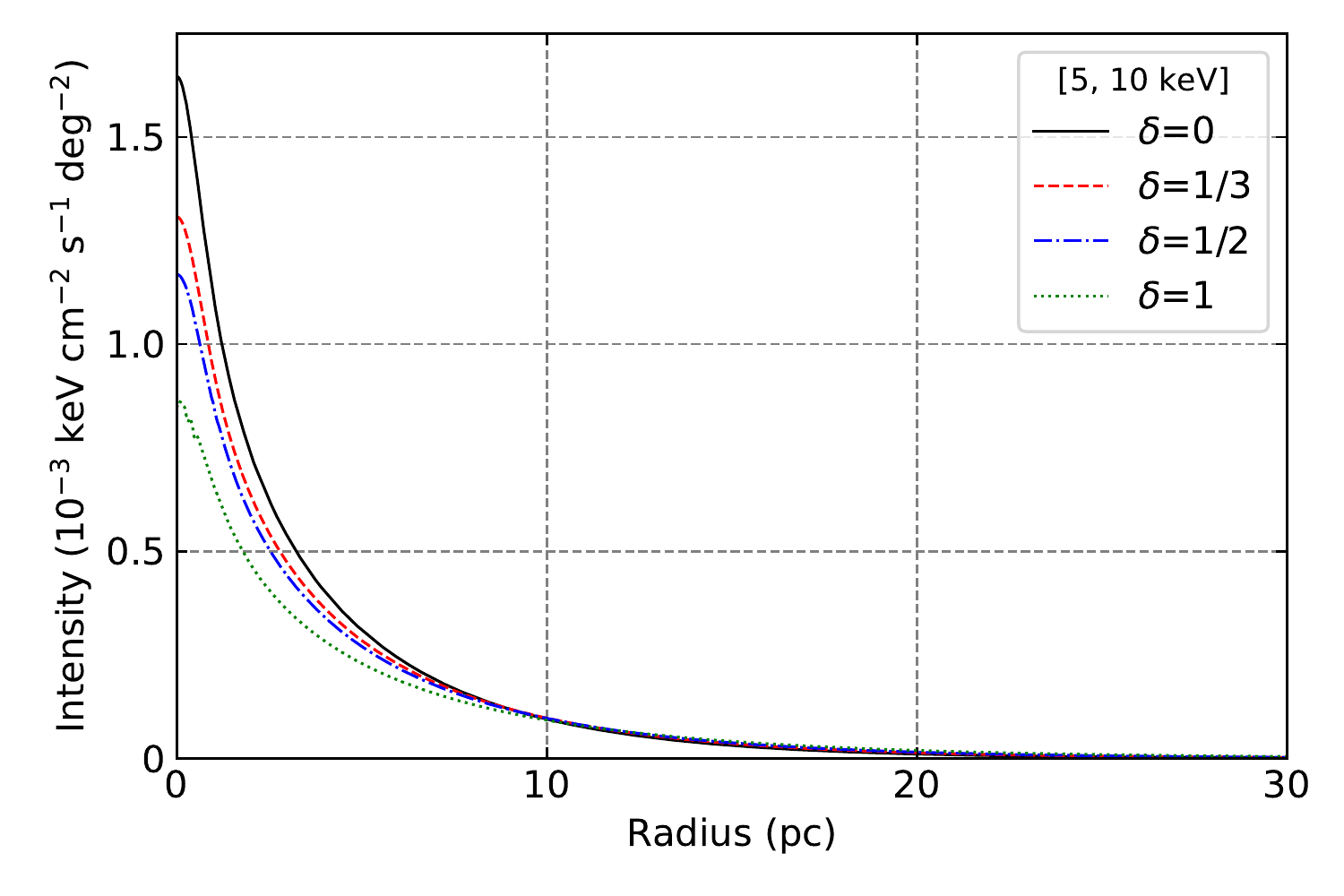}
	\caption{The  X-ray surface brightness profile of the halo around Geminga for different  $\delta$, which is defined as $D(E)\propto E^{\delta}$.}
	\label{fig:5}
\end{figure*}
\bsp	
\label{lastpage}
\end{document}